\documentclass[usenatbib]{mnras}
\usepackage{graphicx}
\usepackage{txfonts,times}

\usepackage{lipsum}
\usepackage{gensymb}
\usepackage{color}

\newcommand{\Msun}{\,{\rm M}_{\odot}}
\newcommand{\kms}{\,km\,s$^{-1}$}

\title[Sloshing in A2199]{Simulating nearly edge-on sloshing in the galaxy cluster Abell 2199}
\author[Machado et al.]{R. E. G. Machado$^{1}$\thanks{E-mail: rubensmachado@utfpr.edu.br},
T. F. Lagan\'{a}$^{2}$,
G. S. Souza$^{2}$,
A. Caproni$^{2}$,
A. S. R. Antas$^{2}$, and
\newauthor E. A. Mello-Terencio$^{3}$
\\
$^{1}$Departamento Acad\^emico de F\'isica, Universidade Tecnol\'ogica Federal do Paran\'a, Rua Sete de Setembro 3165, Curitiba, PR, Brazil \\
$^{2}$N\'ucleo de Astrof\'isica, Universidade Cidade de S\~ao Paulo, Galv\~ao Bueno 868, Liberdade, 01506-000, S\~ao Paulo, SP, Brazil\\
$^{3}$Departamento de F\'isica, Universidade Federal do Paran\'a, 81531-980, Curitiba, PR, Brazil}
\date{Accepted 2022 June 27. Received 2022 June 14; in original form 2022 April 05}
\pubyear{2022}

\begin{document}  
\label{firstpage}
\pagerange{\pageref{firstpage}--\pageref{lastpage}}
\maketitle

\begin{abstract}
Off-axis collisions between galaxy clusters may induce the phenomenon of sloshing, causing dense gas to be dragged from the cool core of a cluster, resulting in a spiral of enhanced X-ray emission. Abell~2199 displays signatures of sloshing in its core and it is possible that the orbital plane of the collision is seen nearly edge-on.
We aim to evaluate whether the features of Abell~2199 can be explained by a sloshing spiral seen under a large inclination angle.
To address this, we perform tailored hydrodynamical $N$-body simulations of a non-frontal collision with a galaxy group of $M_{200}=1.6\times10^{13}\,{\rm M_{\odot}}$.
We obtain a suitable scenario in which the group passed by the main cluster core 0.8\,Gyr ago, with a pericentric separation of 292\,kpc. Good agreement is obtained from the temperature maps as well as the residuals from a $\beta$-model fit to the simulated X-ray emission. We find that under an inclination of $i=70\degree$ the simulation results remain consistent with the observations.
\end{abstract}


\section{Introduction}

Galaxy clusters are the most recent objects in the hierarchical scenario of large scale structure formation. Due to mergers and collisions, the gas of the intracluster medium (ICM) often shows signs of recent disturbances such as cold fronts and shocks \citep[e.g.][]{Markevitch2007}.

The phenomenon of gas sloshing \citep{Markevitch2001} arises in the case of an off-axis collision between two clusters. Due to the gravitational perturbation induced by the close passage of a substructure, the cool gas that was in equilibrium in the main cluster core is driven out. Depending on the parameters of the clusters and of the collision, this can in general produce a relatively large spiral feature, manifested as an excess in X-ray surface brightness. This spiral of cool dense gas stems from the core and may commonly reach a few 100\,kpc in extent, or even as much as $\sim$700\,kpc as revealed by deep X-ray observations in the Perseus Cluster \citep{Walker2018, Walker2022} or Abell 2142 \citep{Rossetti2013}.

These spiral features have been observed in the ICM of several galaxy clusters. One of the most notable is that of Perseus \citep[][]{Churazov2003}. Other examples of clusters with signs of sloshing include: Abell~85 \citep{Ichinohe2019}, Abell~2029 \citep*{Clarke2004, PaternoMahler2013}, Abell~1644 \citep{Johnson2010}, Abell~2052 \citep{Blanton2009, Blanton2011}, Abell~2142 \citep{Owers2011, Rossetti2013, Liu2018}, Abell~496 \citep{Roediger2012b, Ghizzardi2014}, RXJ2014.8-2430 \citep{Walker2014}, Abell~1835 \citep{Ueda2017}, Abell~1763 \citep{Douglass2018}, among others \citep*[see e.g.][]{Owers2009, Ghizzardi2010}. Spiral-like structures have also been observed in groups \citep[e.g.][]{Randall2009, Lal2013, Gastaldello2013}. From a sample of nearby galaxy clusters, \cite{Lagana2010} found that half of them have signs of spiral-like structures.

Hydrodynamical simulations have been widely used to understand the sloshing mechanism as well as to explore the parameter space of cluster mergers in general \cite[e.g.][]{AscasibarMarkevitch2006, ZuHone2010, ZuHone2011}. Idealized binary collisions have proven extremely useful in understanding the dynamical history of merging clusters. Such simulations are vital to help interpret the observed configurations and have been employed often to model gas sloshing in individual clusters, such as: Virgo \citep{Roediger2011}, Abell~496  \citep{Roediger2012b}, RXJ1347.5-1145 \citep{Johnson2012}, Abell~2052 \citep{Machado2015}, Abell~1644 \citep{Doubrawa2020, MonteiroOliveira2020}, Perseus \citep{Walker2018}, Fornax \citep{Sheardown2018}, among others.

A few examples exist in the literature of merging clusters interpreted as seen nearly along the line of sight \citep{Dupke2007}, including ones with gas sloshing \citep{Ueda2019}. The most favorable geometry for viewing the sloshing spiral is the one where the orbital plane of the clusters coincides with the plane of the sky, but even in this optimal configuration, the gas morphology is not always trivial to interpret without the aid of simulations. In the case of large inclinations, the projected morphology is even less intuitive, since the spiral perturbation is not truly confined to a thin layer of gas. Although the sloshing signature is not fully suppressed under large inclinations \citep{Roediger2011}, it tends to become substantially weaker and to be characterized by arcs rather than spiral arms. The detailed morphology will depend on the specific properties of the clusters as well as on the age of the collision, but in the general case it would be difficult to predict a priori what the typical appearance of an arbitrarily inclined sloshing spiral should be. Thus dedicated simulations are needed in order to argue for a specific collision scenario.

An example of a cluster that presents signatures of mergers and/or collisions -- possibly with large inclination -- is Abell\,2199 (A2199), a relatively nearby rich cluster of galaxies with a redshift of $0.030151\pm 0.000230$ \citep{2001AJ....122.2858O}. Its inferred temperature profile indicates a cooler central region ($< 100$\,kpc) in relation to its outskirts, where the temperature remains aproximately constant ($\sim$4\,keV; e.g., \citealt{1998MNRAS.298..416P, 2002MNRAS.336..299J}). The cool core of A2199 harbours a massive cD galaxy, NGC\,6166, which has a radio source labelled as 3C\,338 (e.g., \citealt{1983ApJ...271..575B}), from which emanate a jet and a counter-jet that interact with the surrounding material (e.g., \citealt{1998ApJ...493...73O, Nulsen2013}). Observations at X-ray energies have revealed the presence of low-density cavities, shocks, as well as filamentary structures in the core region of A2199 (e.g., \citealt{1998ApJ...493...73O, 2002MNRAS.336..299J, 2004ApJ...607..800B, 2006MNRAS.372...21A, 2006MNRAS.371L..65S, Nulsen2013}). Moreover, \citet{Nulsen2013} proposed that asymmetries seen at larger scales (excess X-ray emission detected 200$''$ away southwest of the center of A2199, and a low-entropy region 50$''$ north from the same center) could be signatures of gas sloshing induced by the passage of a merging subcluster 400\,Myr ago. Indeed, A2199 forms a supercluster with neighboring clusters and groups of galaxies that are likely falling into it \citep{2001ApJ...555..558R}, corroborating such sloshing possibility for this system.

In this paper we aim to reconstruct the dynamical history of A2199 by simulating an off-axis collision with a galaxy group. In particular, we wish to evaluate whether the observed spiral-like feature is consistent with the scenario of a collision seen almost edge-on, or at least under a large inclination angle.

The paper is divided as follows. In Section~\ref{sec:methods} we present the methods for observational analysis and the simulation setup. In Section~\ref{sec:results} we present the results of the X-ray observational data, the identification of structures in the optical and the simulation results. Summary and conclusions are given in Section~\ref{sec:conclusions}. This paper adopts a flat $\Lambda$CDM cosmology with [$\Omega_{\rm M}, \Omega_{\Lambda}, H_{0}$] = [0.27, 0.73, 70 km\,s$^{-1}$\,Mpc$^{-1}$].

\section{Methods}
\label{sec:methods}

\subsection{X-ray data analysis}

\subsubsection{Chandra}
We use Chandra data (ObsID 10748 and 10803) for spectral analysis, following the steps in {\sc ClusterPyXT}\footnote{https://github.com/bcalden/ClusterPyXT} \citep{Clusteryxt19} that automates the creation of data products. Briefly, the user provides key information such as ObsID, the galactic hydrogen column density along the line of site ($n_{H}$), redshift, among others and the software proceeds to download the data and merge the observations and corresponding backgrounds in the [0.7--8.0] keV energy range. Then, the user should provide a list of point source regions to be removed and the pipeline proceeds to filter high energy flares. 
The next stage calculates the bins using adaptive circular binning algorithm. This process creates the regions for each spectral fit, and prepares the script for the final stage. The final step of the procedure consists on the spectral fitting for each pixel to generate temperature, pressure and entropy maps. The temperature map with high resolution results in a spectral temperature fit for every pixel in the X-ray image.

\subsubsection{XMM-\textit{Newton}}

We use XMM-\textit{Newton} archival data for A2199 (ObsID 0723801101) and the data reduction was done with Science Analysis System (SAS, version 18.0.0) and calibrated files update to 2020. The steps were described in detail in \citet{Lagana08,Durret10,Durret11,Lagana13,Lagana19} and are summarized below:
\begin{itemize} 
\item To filter background flares with a 2-$\sigma$ clipping procedure in the [10--14] keV energy band light curve;
\item To obtain a normalization parameter to be applied in the spectral fits matching a background spectrum in an outer annulus of the observation in the [10--12] keV energy band with the background spectrum obtained with the blank sky \citep{Read03}  in the same energy band and region;
\item Exclude point sources from the event files applying a visual inspection and confirming them in the High Energy Catalogue 2XMMi Source.
\end{itemize}

The spectral analysis was restricted to the energy range [0.7--7.0] keV and to avoid any influence from Al and Si instrumental lines we also exclude the [1.2--1.9] keV energy band. All spectra are re-binned to ensure at least 15 counts in each energy bin. We adopted WABS(APEC), an absorbed thermal plasma model and abundance ratios from \citet{Asplund09}.

For both Chandra and XMM-\textit{Newton} data, the fitting procedure was done in XSPEC version 12.9.1, with redshift and $n_{H}$ values fixed at $z=0.0306$ and $n_{H} = 8.92 \times 10^{20}\,\rm atoms \, cm^{-2} $.

\subsection{Caustic technique}
The caustic technique \citep{Diaferio97,Diaferio99} is designed to estimate the mass profile of galaxy clusters to radii well beyond the virial radius, and two by-products of this technique are the identification of the cluster members and the identification of the cluster substructures. 

We first select all spectroscopically identified galaxies in SDSS-DR16 (Sloan Digital Sky Survey - Data Release 16),  
within a radius of approximately $3 \times R_{200}$ 
\citep[where $R_{200} = 1.526\; \rm Mpc$;][]{Piffaretti11},
and $r$-band magnitudes $m_{r} \leq 17.78$,
which is the survey spectroscopic completeness limit, that at the cluster redshift correspond to  all galaxies with $M_{r} < -17.81$.

To identify among these selected galaxies the ones that are gravitationally bound to the cluster, we used the CausticMass code \citep{Gifford13,Gifford-Miller13}. This approach uses the positions and velocities of galaxies and projects them onto the  $R-v$ diagram that shows the rest-frame velocity of galaxies relative to the cluster center as a function of their clustercentric distance. The caustic technique assumes only spherical symmetry and so the system does not need to be in dynamic equilibrium. In this way it is possible to use this technique both in the central regions and in the outermost radii, where other techniques cannot be applied \citep{Yu15}. 

\begin{figure}
\center
\includegraphics[width=0.8\columnwidth]{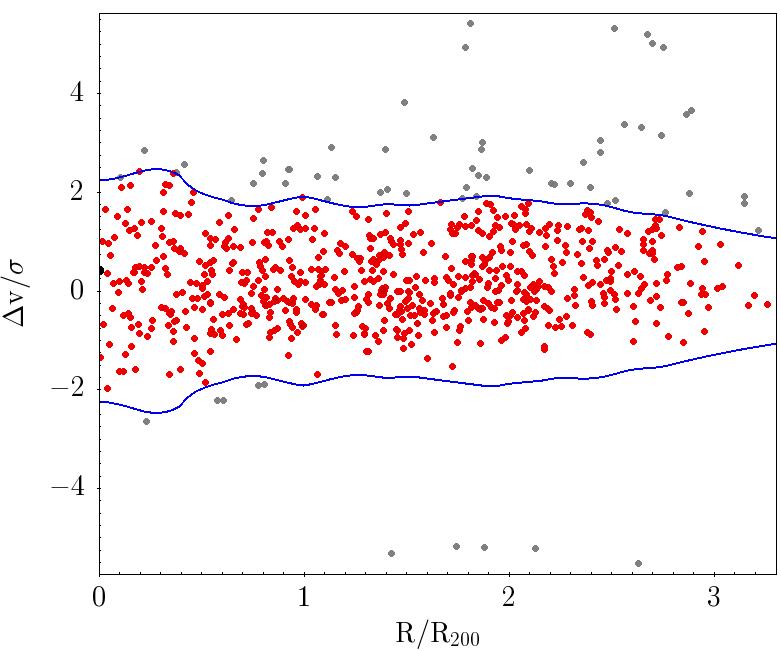}
\caption[]{Phase-space diagram for A2199: the peculiar velocity 
as a function of the clustercentric distance. The velocity axis is normalized by the velocity dispersion of the cluster along the line-of-sight ($\sigma$), while the radial distance is given in units of $R_{200}$. The solid curves show the caustics, and the red points in between are the 674 cluster members.}
\label{figCaustic}
\end{figure}

As shown in Fig.~\ref{figCaustic}, galaxies are distributed in this phase space in a region that looks like a trumpet shape \citep{vanHaarlem93}. The caustics define the limits in the phase space where the speed of galaxies is less than the escape velocity, defining 674 cluster members. 

\begin{figure}
\center
\includegraphics[width=0.85\columnwidth]{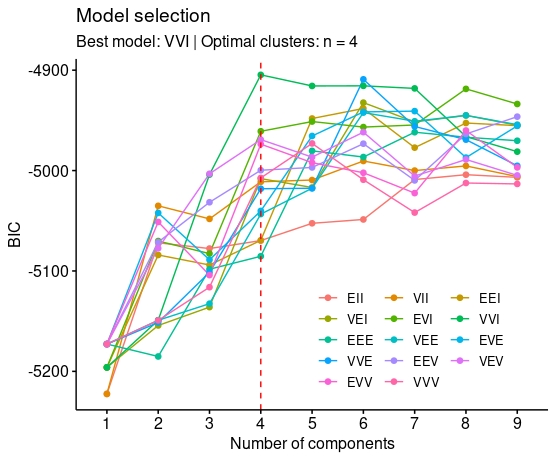}
\caption[]{BIC analysis for the different models fitted by {\sc mclust}. The three-component model VVI with four components is the one to better fit the data.}
\label{figBIC}
\end{figure}

\subsection{Substructure Identification}
  
After applying the caustic technique to define the member galaxies of A2199,  we use the R package {\sc mclust} \citep{Fraley06}. This method uses a Gaussian Mixture Model (GMM) along with hierarchical clustering to group a set of data points into subclusters. The main aim is to creat subclusters that are coherent internally and  clearly different from the other subgroups \citep{Scrucca16}. 
 
These GMMs consist of probabilistic models that fit a finite number of gaussian distributions that have an unknown mean and covariance with respect to the dataset under study. The Expectation-Maximisation statistic is used, which initially considers the parameters that describe the Gaussian as fixed and calculates for each point the probability of belonging to each subcluster. Then, the probability of each point belonging to a given cluster is fixed and the parameters that describe the Gaussian are weighted by the sum of the probabilities of each point belonging to the clusters \citep{Ana20}. These two iterations continue until converging.

\begin{figure}
\center
\includegraphics[scale=0.45]{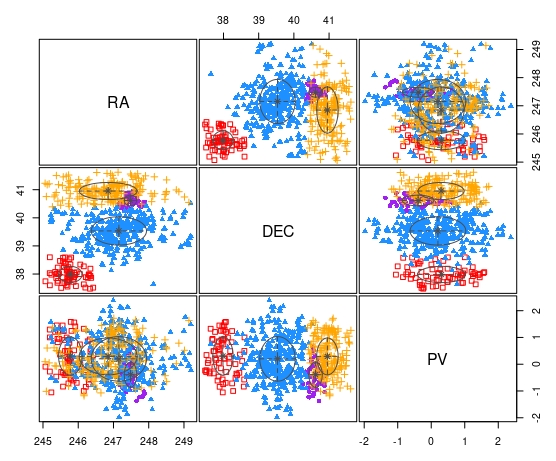}
\caption[]{Class membership using  {\sc mclust} in sky coordinates (in degrees) and peculiar velocity measured along the line of sight (PV) space. The velocity axis is normalized by the velocity dispersion of the cluster along the line-of-sight ($\sigma$). The subgroup represented by blue points is the main one with X-ray counterpart. The points represented by other colours (red, yellow, and purple) are subgroups identified by {\sc mclust} for A2199.}
\label{figMclust}
\end{figure}

\begin{figure}
\center
\includegraphics[scale=0.2]{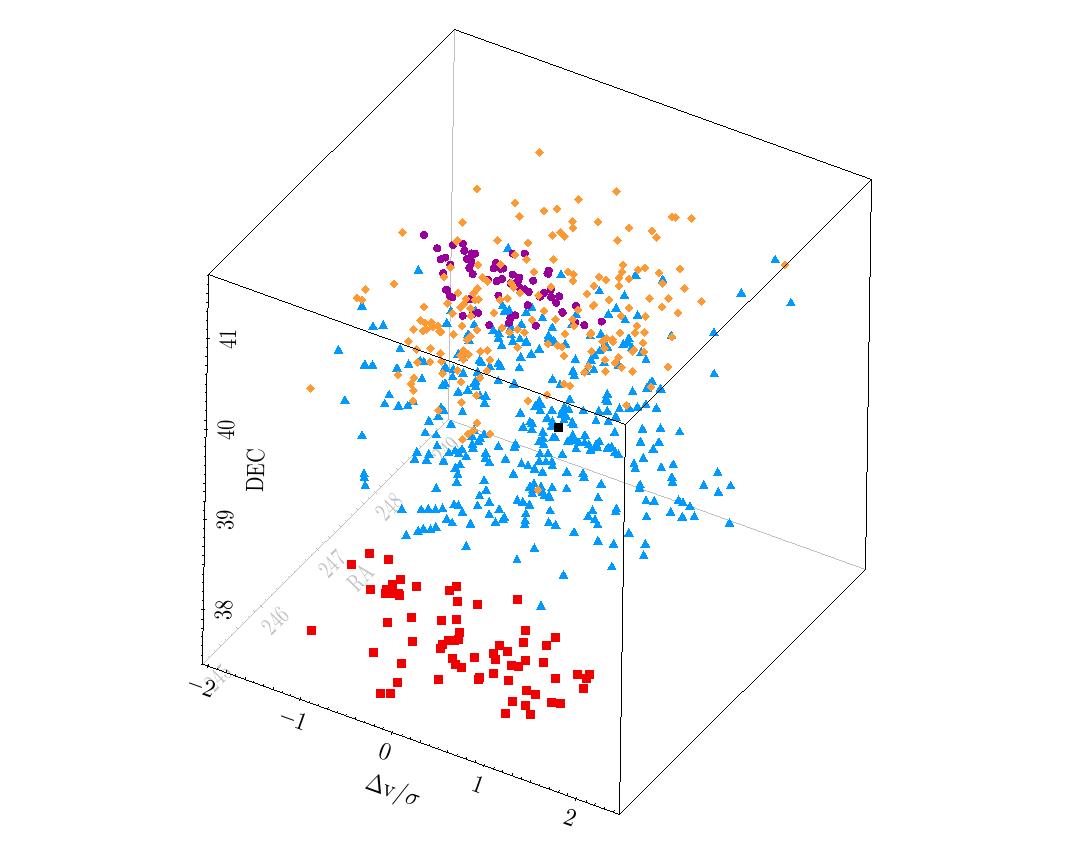}
\caption[]{Distribution of galaxy subgroups identified by {\sc mclust} in 3D, with colors accordingly.}
\label{fig3D}
\end{figure}

In summary, {\sc mclust} calculates the log-likelihood for each iteration and then computes the most effective approximation to estimate the mixtures by varying the shape, orientation, and volume, of multidimensional Gaussians. We perform the {\sc mclust} in its 3-dimentional mode, having as input the radial velocity and sky coordinates (RA and Dec). According to \citet{Ana20}, the most credible model for the number of groups is the one with the highest Bayesian Information Criterion \citep[BIC;][]{KassRaftery1995}, as shown in Fig.~\ref{figBIC}.

Thus, we assume four optical substrucutres (the main cluster and three subgroups), spatially distributed as shown in Figs.~\ref{figMclust} and~\ref{fig3D}, to proceed with our analysis. The results are described in Section~ \ref{sec:optical}.

\subsection{Simulation setup}

We wish to model the current state of A2199 by simulating an off-centre binary collision between a main cluster and a subscluster. Each structure is represented by a spherically symmetric halo comprising dark matter (DM) particles and gas particles. The methods for creating initial conditions are described in greater detail in \cite{Machado2013} or \cite{Ruggiero2017}, for example, and here we give a brief summary of the main features.

The dark matter haloes are represented by a \cite{Hernquist1990} profile:
\begin{equation}
\rho_{\rm h}(r) = \frac{M_{\rm h}}{2 \pi} ~ \frac{r_{\rm h}}{r~(r+a_{\rm h})^{3}} \, ,
\end{equation}
where $M_{\rm h}$ is the total dark matter mass, and $a_{\rm h}$ is a scale length. The gas follows a \cite{Dehnen1993} density profile, with gas mass $M_{\rm g}$ and scale lenght $a_{\rm g}$:
\begin{equation}
\rho_{\rm g}(r) = \frac{(3-\gamma)~M_{\rm g}}{4\pi} ~ \frac{r_{\rm g}}{r^{\gamma}(r+a_{\rm g})^{4-\gamma}} \, .
\end{equation}
The choice of $\gamma=1$ leads to a cool core, which is required in the main cluster, while $\gamma=0$ produces a non-cool core in the smaller subcluster. Once the density profiles are set, imposing hydrostatic equilibrium determines the gas temperatures. Numerical realizations of these profiles are generated according to the techniques outlined in \cite{Kazantzidis2006}. The positions of the dark matter particles and of the gas particles are obtained by uniformly sampling the inverse of the cummulative mass function. The velocities of the dark matter particles are obtained by integrating Eddington's formula numerically. This method has the advantage of not assuming a local Maxwellian distribution of velocities \citep{Kazantzidis2004}. Additional details are described in \cite{Machado2013}. To create initial conditions for this work, we employed a version\footnote{\href{https://github.com/elvismello/clustep}{https://github.com/elvismello/clustep}} of the {\sc clustep} code \citep{Ruggiero2017} ported to Python~3.

To set up the initial conditions, the structural parameters of the clusters need to be chosen such as to meet certain observational constraints. The lack of knowledge of the past state of the clusters is part of the challenge when attempting to model their history. In the case of A2199, the observational data does not suggest a particularly violent recent collision, meaning that the overall mass distribution should not have been dramatically altered. Thus it seems reasonable to assume that the current features of the cluster might be a good first approximation for the initial conditions -- namely its mass profile and its temperature profile.

\begin{figure}
\includegraphics{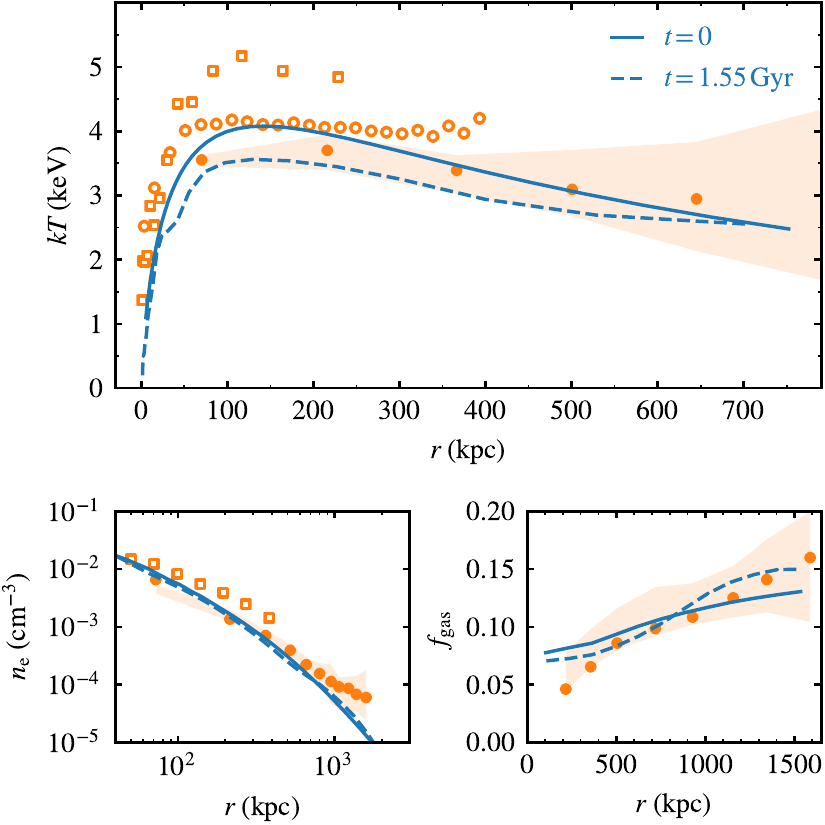}
\caption[]{Profiles of temperature, electron number density and gas fraction. The solid/dashed blue lines represent the initial conditions and the best moment of the simulation. The orange symbols represent observational constraints: solid circles and shaded areas from \cite{Mirakhor2020}; open squares from \cite{Nulsen2013}; and open circles from this work.}
\label{002b}
\end{figure}

\begin{table}
\caption{Properties of the initial conditions for the main cluster.}
\label{tab:ic}
\begin{center}
\begin{tabular}{l c}
\hline
$M_{200}$          & $3.0 \times 10^{14}\Msun$ \\
$M_{500}$          & $2.5 \times 10^{14}\Msun$ \\
$r_{200}$          & 1.4\,Mpc \\
$r_{500}$          & 1.0\,Mpc \\
concentration    & 4.4 \\
$f_{\rm gas}$ at $r_{200}$  & 0.13 \\
${kT}~(r < 50\,{\rm kpc})$ &  2.5\,keV \\
\hline
\end{tabular}
\end{center}
\end{table}

The main observational constraints that need to be approximately satisfied are as follows. We use as reference the virial mass $M_{200} = ( 3.10 \pm 0.25 ) \times 10^{14}\Msun$ or $M_{500} = ( 2.41 \pm 0.15 ) \times 10^{14}\Msun$ from \cite{Mirakhor2020}; or still $M_{200} = ( 3.16 \pm 0.48 ) \times 10^{14}\Msun$ from \cite{Lee2015}. These masses are consistent with the scaling relations from \cite{Piffaretti2011} and \cite{Lovisari2015}. Likewise, the virial radius should be $r_{200} = 1.60$\,Mpc, and the gas fraction at $r_{200}$ should be $f_{\rm gas} = 0.16$ \citep[both from][]{Mirakhor2020}. Given this $M_{200}$ at redshift $z=0.03$, the concentration of such a halo ought to be approximately $c=4.3$ according to \cite{Duffy2008}. From the Chandra temperature maps, we see that the mean temperature within $\sim$50\,kpc is below 3.0\,keV.

Based on these target constraints, the initial conditions for the main cluster were created with total mass of $5.3 \times 10^{14}\Msun$, total baryon fraction of 0.12 and scale lengths $a_{\rm g}=650$\,kpc and $a_{\rm h}=420$\,kpc. These choices of parameters led to the virial mass, virial radius, concentration and temperature listed in Table~\ref{tab:ic}, which are very comparable to the target observational constraints. The temperature profile of the initial condition is shown in the top panel of Fig.~\ref{002b}, in comparison to the observational data (orange points). The results of the simulation will be discussed in detail in Section~\ref{sec:sim}, but here we already present the temperature profile measured at a later time in the simulation, in order to show that the temperature structure was sufficiently robust for our purposes. In other words, setting the currently observed temperature profile at $t=0$ proved to be a good approximation, since it varied only by about 0.5\,keV. It should be noted that the departures of 0.5\,keV seen in the dashed line of Fig.~\ref{002b} correspond to the actual collision simulation, including the perturber. But if the main cluster is relaxed in isolation, its azimuthally averaged temperature profile is quite similar. In what follows, we will be mostly concerned with deviations from an azimuthally averaged profile only in the inner region $r \lesssim 100$\,kpc. Additionally, Fig.~\ref{002b} displays the radial profiles of gas density and of gas fraction, comparing the initial conditions to the observational data points from \cite{Mirakhor2020} and \cite{Nulsen2013}. These quantities also remain sufficiently robust during the simulation, justifying the choices at $t=0$.

For the subcluster, we adopted a mass ratio of $1:20$, meaning that it actually represents a galaxy group with virial mass of approximately $1.6 \times 10^{13}\Msun$, virial radius of $r_{200}=0.5$\,Mpc and corresponding concentration of approximately $c=5.9$.

Having created the two structures with the parameters described above, we finally prepare the collision setup by setting them apart with an initial separation of 2\,Mpc, an impact parameter $b=400$\,kpc, and an initial relative velocity $v_0=-2500$\kms. The simulations were performed with the smoothed particle hydrodynamics (SPH) $N$-body code {\sc Gadget-2} \citep{Springel2005}, using a gravitational softening length of 1\,kpc for all particles, and the evolution was followed for at least 2\,Gyr. The main cluster has $10^6$ gas particles and $10^6$ dark matter particles; the subcluster has proportionally fewer particles such as to keep the same mass resolution of each particle type.

\section{Results}
\label{sec:results}

\subsection{X-ray analysis}
\label{sec:xray}

\begin{figure*}
\centering
\includegraphics[scale=0.5]{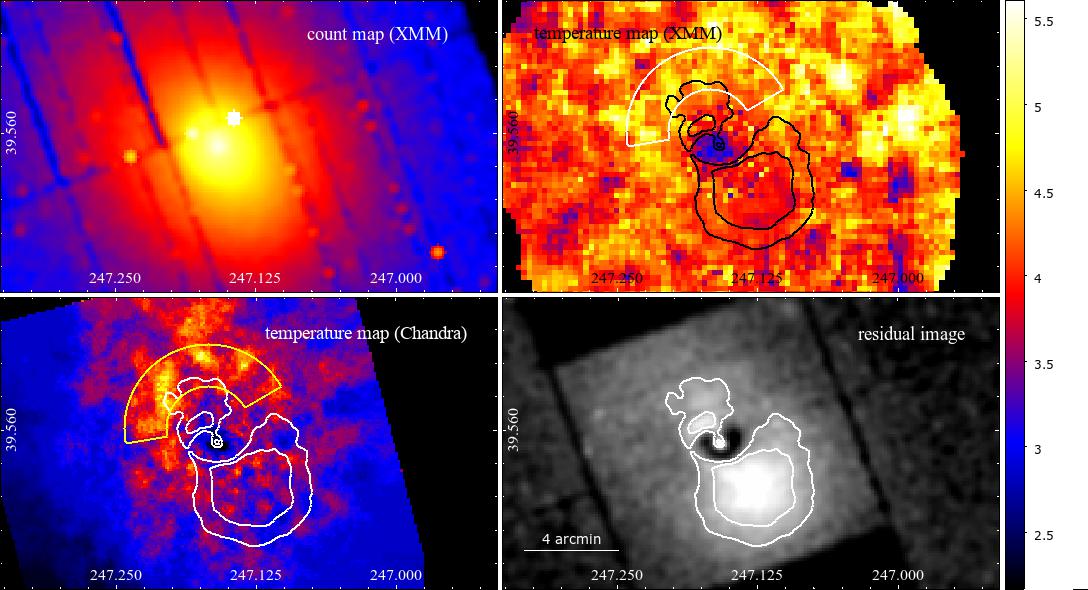}
\caption[]{Top panel: Count and temperature maps from XMM-\textit{Newton}, Lower panel: Chandra temperature map and residual imagem. The white contours represent the sloshed gas and the yellow semi-circular anullus the higher temperature region.
 Color bars indicate temperature values in keV.}
\label{figkTmaps}
\end{figure*}

In this section we present the temperature maps obtained from XMM-\textit{Newton} and Chandra data. Fig.~\ref{figkTmaps} shows XMM-\textit{Newton} and Chandra temperature maps as well as the count map and residual images. From XMM-\textit{Newton} we have an overall view of the temperature distribution showing a cool core region surrounded by an almost isothermal temperature distribution around 4--5 keV and some hotter spots (yellow regions) towards the outer parts of A2199. Due to its spatial resolution, the Chandra temperature map shows a complex structure with a hotter region (highlighted by the yellow semi-circular annulus), 200 arcsec from the center and opposite to the sloshed gas, that is also present in our hydrodynamical simulation results. The cooler central region has a pattern (more clear in the upper right panel of Fig.\ref{005m}) that resemble in morphology the H-$\alpha$ filaments detected in Hubble Space Telescope images, although in a much larger scale. These plumes with temperatures below 1\,keV in the very inner region are evidence of the cooling flow in A2199 that is probably, feeding the central AGN. The residual map clearly show the spiral feature embedded in the global emission of the cluster and its coincidence with regions of lower temperature, corroborating with the sloshing scenario.
 
\subsection{Optical substructures}
\label{sec:optical}
Using membership assignment, the caustic method identified 674 galaxies as cluster members. Then, using the spatial coordinates plus the radial velocity, we can search for optical substructures.  {\sc mclust} analysis found four overdensities within 4.578 Mpc ($3 \times R_{200}$) from the cluster X-ray emission peak: the main cluster, that has an X-ray counterpart and three subgroups. Then, for each subgroup and the corresponding galaxy membership, we use again the caustic method to estimate the characteristic radius ($r_{200}$) and masses ($M_{200}$) summarized in Table \ref{tab:mclust}. The listed values indicate that the main cluster is more than 40 per cent more massive than the main subgroup (the orange one) and an order of magnitude more massive than the smaller one. 

\begin{table}
\caption{Main properties of A2199 and it subgroups according to membership assignment. The colors stated in the first column follow Figs.\ref{figMclust} and \ref{fig3D}, and the second column gives the number of member galaxies of each subgroup.} 
\label{tab:mclust}
\begin{center}
\begin{tabular}{cccc}
\hline
cluster & $N_{\rm memb}$ & $r_{200}$  & $M_{200}$   \\
        &  & (Mpc) &   $(10^{14} {\rm M}_{\odot})$ \\
\hline
Total 			& 674  	& 1.16 	&   3.66 \\
\hline
Main (blue)       	 & 348      & 1.03        & 2.57 \\
Subgroup 1 (orange)  & 193      & 0.85        & 1.47  \\
Subgroup 2 (red)   	 & 68       & 0.76        & 1.05 \\
Subgroup 3 (purple)  & 65       & 0.52        & 0.35 \\
\hline
\end{tabular}
\end{center}
\end{table}

\subsection{Simulation results}
\label{sec:sim}

In this section, we present the results from the $N$-body hydrodynamical simulations. We performed a set of simulations aiming at reproducing several observed features of A2199. The main purpose of this simulation study was to evaluate whether the spiral-like morphology in the cluster core could be explained by a sloshing event seen under a high inclination angle. With this in mind, we report mainly on one specific simulation -- model A -- to which we sometimes refer as the `best model' with its `best instant', but bearing in mind that this is to be understood as one representative model within a family of similarly adequate models.

The bulk structural parameters of the clusters are similar to the observed ones, by construction. Apart from those, the other quantities that the simulation aims to reproduce are: (i) the overall temperature structure in the cluster core, and (ii) the detailed spiral-like morphology of the residuals in the cluster core. These requirements must be met simultaneously, at least within an acceptable approximation. The map of residuals (i.e. simulated X-ray emission subtracted from a fitted $\beta$ model) was the most challenging feature to recover with accuracy.

\subsubsection{The orbit of the perturber}

The substructures identified in Section~\ref{sec:optical} are located roughly 2\,Mpc away from the cluster core. By trial and error, we found best results were obtained when the final position of the perturber is towards the northeast. Other configurations proved difficult to reconcile with the orientation of the spiral-like feature, as we will see in the following. We used in the simulation (model A) a substructure with virial mass $1.6 \times 10^{13}\Msun$, meaning a mass ratio of only $1:20$. This is consistent with the expectation of a mild collision, since the observed spiral-like feature is rather small and subtle.

\begin{figure}
\includegraphics{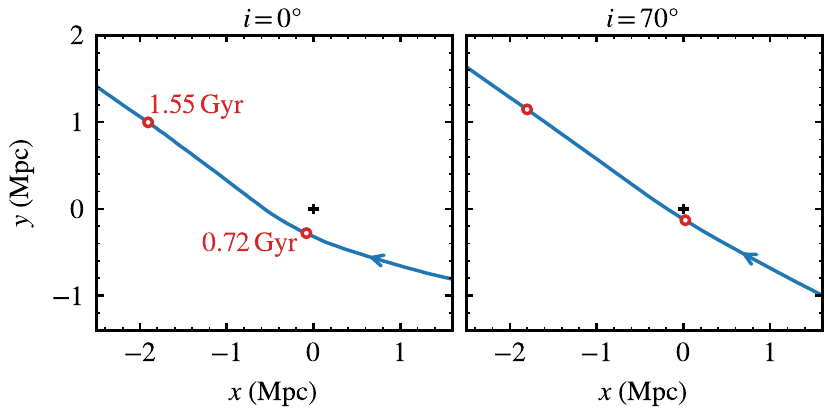}
\caption[]{The orbit of the subcluster (blue line) with respect to the centre of the main cluster (black cross at the origin), shown face-on (left) and projected under an inclination (right). The red circles mark the moment of pericentric passage (0.72\,Gyr), and the best time (1.55\,Gyr) of the simulation.}
\label{003a}
\end{figure}

The orbit of the perturber in model A is shown in Fig.~\ref{003a}. The pericentric passage occurs at $t=0.72$\,Gyr, at which time the separation between the cluster centres is 292\,kpc. Approximately 0.83\,Gyr after the pericentric passage (at $t=1.55$\,Gyr), the so-called best moment of the simulation is reached.

The trajectory in Fig.~\ref{003a} is seen both face-on and under an inclination of $i=70\degree$. We will argue that this considerable inclination is still consistent with the observational constraints. The rotation of an angle $i$ is applied around an axis defined by the line connecting the two structures at the best moment. This is why the pericentric distance appears shortened under projection in the second frame of Fig.~\ref{003a}, but the separation between the structures does not. After that, another rotation is applied, now around the axis defined by the line of sight, merely to adjust the position angle in the plane of the sky. This is done in an attempt to match the orientation of both the spiral morphology and of the putative subcluster's current position.

By the end, the simulated galaxy group has been substantially stripped of most of its gas content, such that its X-ray emission would be nearly undetectable. This gas stripping is expected and it is seen in simulations when a group or cluster passes too close to the core of another \cite[e.g.][]{Doubrawa2020, MonteiroOliveira2020}. Thus at $t=1.55$\,Gyr, the simulated group is located approximately 2.1\,Mpc from the center of the main cluster, on the northeast quadrant. The somewhat high relative initial velocity of $v_0=-2500$\kms was needed for the group to reach the desired separation in the adequate moment.

\begin{figure*}
\includegraphics[width=0.78\textwidth]{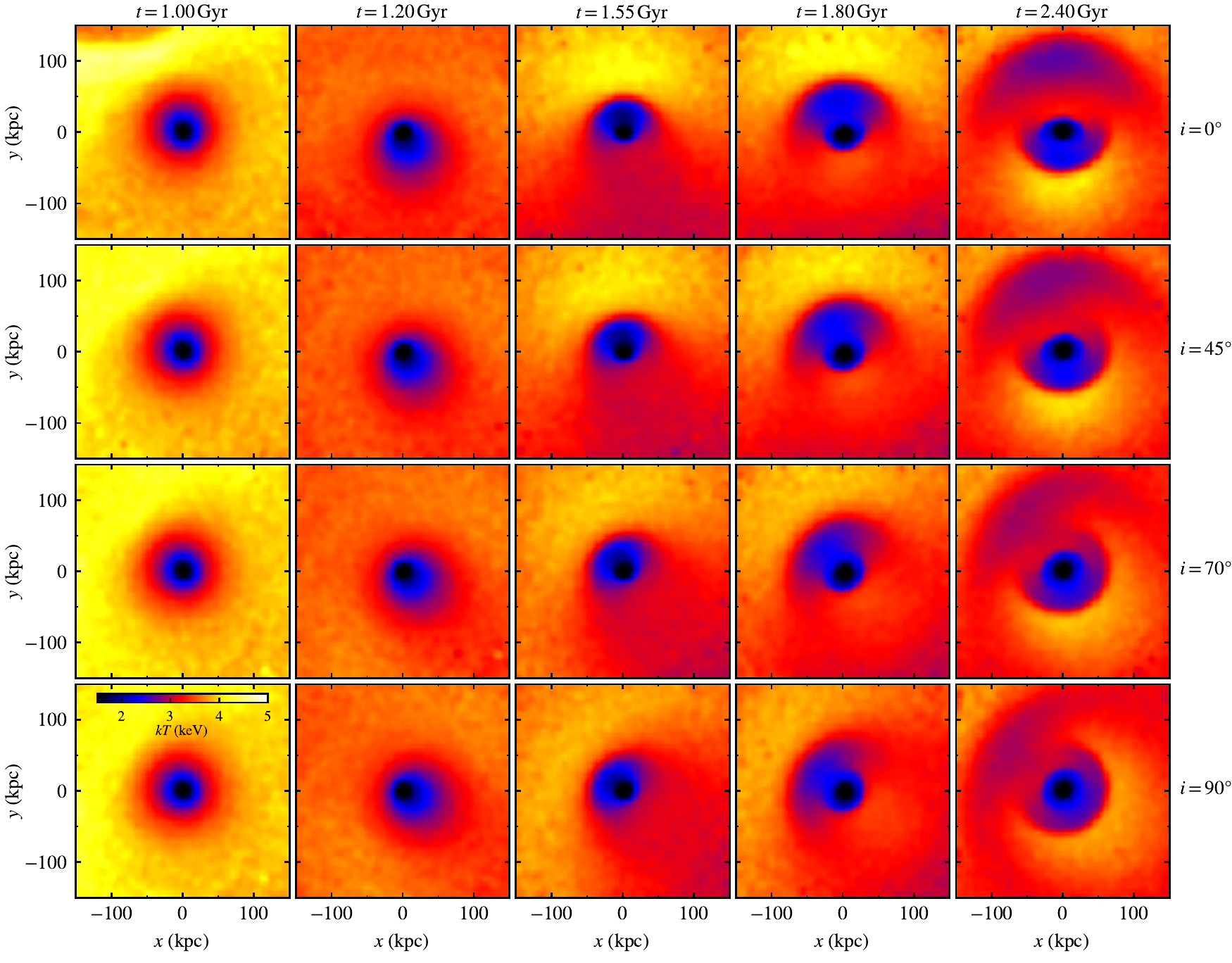}
\caption[]{Emission-weighted temperature maps for the best model, at different times (columns) and under different inclinations (rows).}
\label{001a}
\end{figure*}

\begin{figure*}
\includegraphics[width=0.78\textwidth]{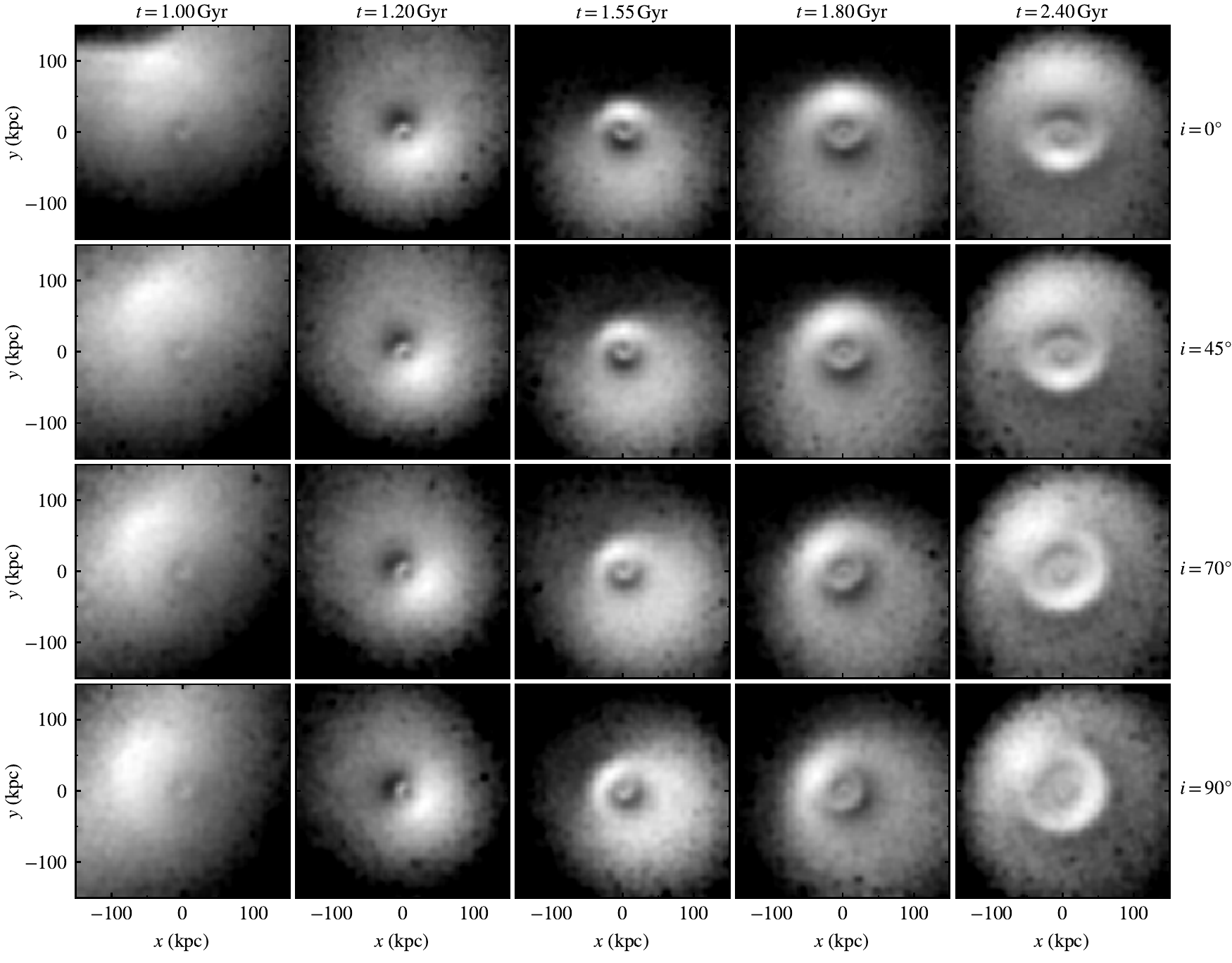}
\caption[]{Residuals between the simulated X-ray emission and a $\beta$-model fit, at different times (columns) and under different inclinations (rows).}
\label{002a}
\end{figure*}

\subsubsection{The spiral-like morphology}

\begin{figure}
\includegraphics[width=\columnwidth]{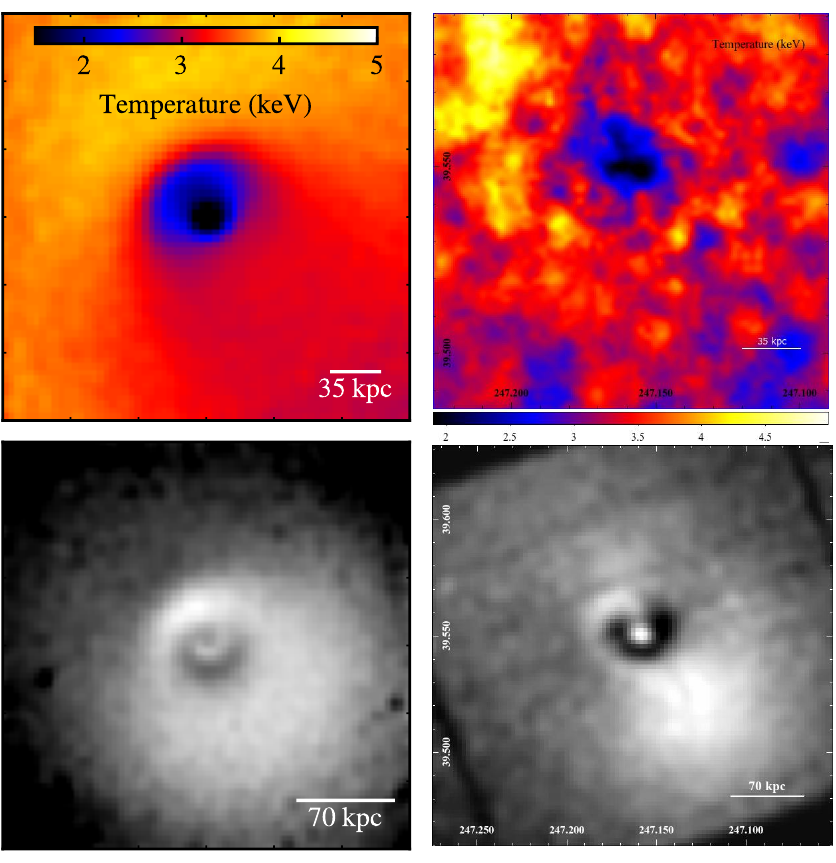}
\caption[]{Comparison to observations for the best model at $t=1.55$\,Gyr with $i=70\degree$: temperature (top) and X-ray residuals (bottom). The simulation results (left) and the observational results from Chandra (top right) XMM-\textit{Newton} (bottom right) are approximately on the same scale.}
\label{005m}
\end{figure}

\begin{figure*}
\includegraphics{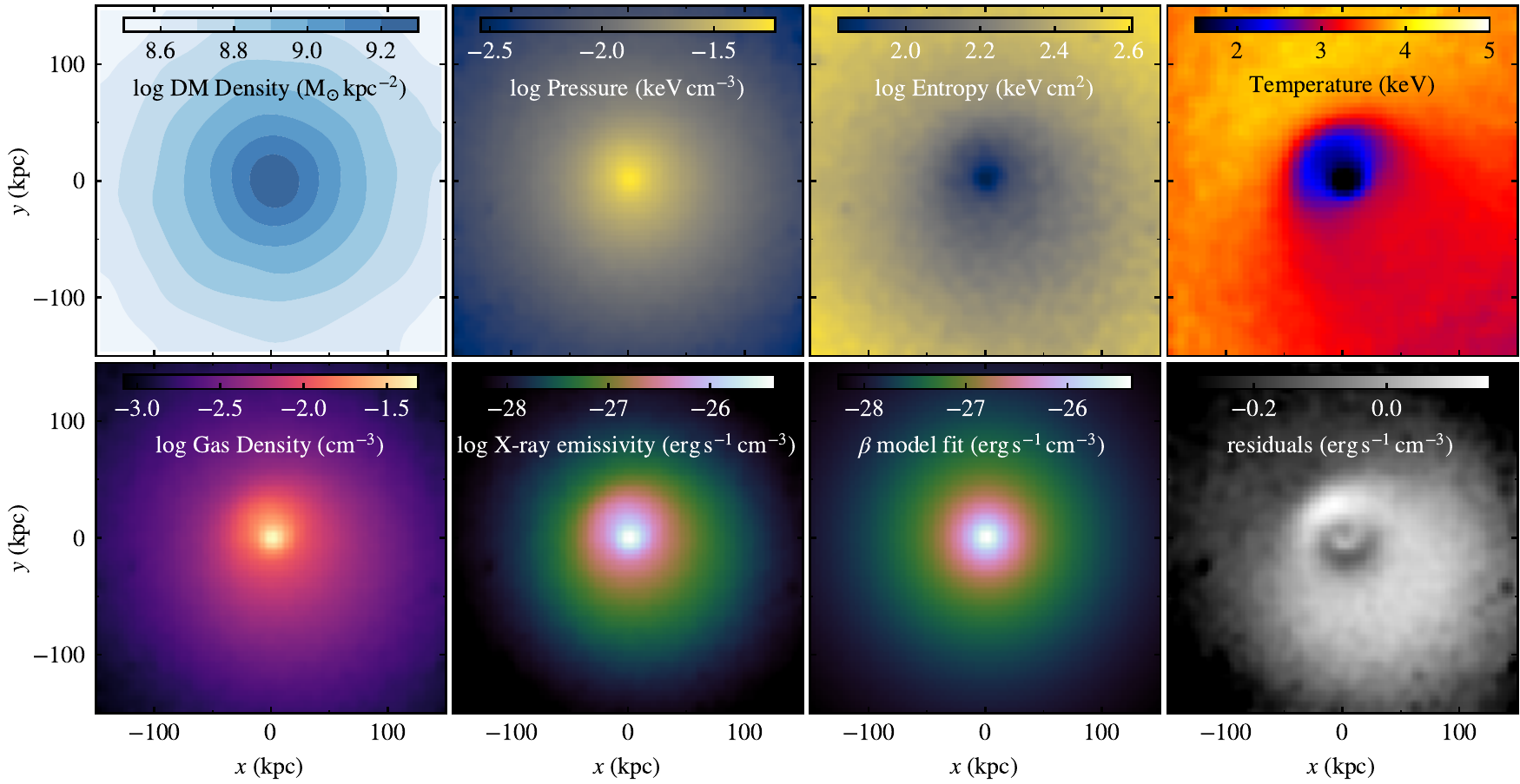}
\caption[]{The best model at $t=1.55$\,Gyr with $i=70\degree$. The quantities shown are: projected dark matter surface density, pressure, entropy, temperature, electron number density, X-ray emission, $\beta$-model fit, and residuals.}
\label{004a}
\end{figure*}

Having established the mechanical aspects of the orbit, we will now look at the consequences of this encounter on the gas in the main cluster core.

In Fig.~\ref{001a} we show the emission-weighted projected temperature maps for model A. These are presented as a function of time to display the evolution of the system. The central column corresponds to the best moment ($t=1.55$\,Gyr); the other columns show selected times that are not equally spaced but were chosen to illustrate relevant features in the evolution. After the central passage, the asymmetry begins to develop. Simulated spirals often extend to a few 100\,kpc \citep[e.g.]{Ascasibar2006, ZuHone2010}, but that is because they are often focused on combinations of parameters that lead to a pronounced effect. In the particular case we are studying, rather than exhibiting a well developed spiral, the sloshing here is modest, mainly due to the low mass of the perturber. On the other hand, depending on the combination of parameters, it is not uncommon for sloshing simulations to display mild asymmetries in the temperature maps, such as features that look like plumes or detached cold fronts, particularly at early times \citep[e.g.][]{Ascasibar2006}.

The first row of Fig.~\ref{001a} shows the encounter seen face-on ($i=0\degree$) and the other rows are projected under certain inclinations, which leads to the issue of projection effects. The trajectory of the substructure does define an orbital plane. This plane may coincide with the plane of the sky, in which case the morphology of a general sloshing spiral is more readily understood. If there is a large inclination $i$ between the plane of the orbit and the plane of the sky, then the morphology of the sloshing spiral becomes quite non-intuitive. Although the perturbation is triggered by an object that moves on a plane, the disturbances in the gas are not confined to the plane, nor even to a thin layer above and below the plane. Because of this tridimensionality, when projecting the temperature along an arbitrarily inclined line of sight, the contributions of the different regions of the gas will result in morphologies that are not obvious to interpret based on one frame alone. In fact, a simulated sloshing spiral, even when well developed and large, may exhibit unintuitive morphologies if projected along an axis not perpendicular to the orbital plane \cite[e.g.][]{Ascasibar2006, Roediger2011}. The sloshing in the model we are analysing here is particularly weak, and the spiral pattern is hardly discernible in the temperature maps, without the aid of the X-ray residuals. These considerations about projection effects on gas properties are also relevant when analysing shock fronts in cluster mergers \cite[e.g.][]{Machado2013, Lagana2019, Chadayammuri2021}.

Concerning the choice of best moment, we see in the temperature maps of Fig.~\ref{001a} that the vicinity of $t=1.55$\,Gyr is in principle favoured because the observations require somewhat higher temperatures ($kT>4$\,keV) on the northeast side in contrast to the southwest. At earlier times, this contrast is not yet established; at later times, there is too much extentend structure around the cool core. Concerning the choice of inclination, none could be strongly ruled out based solely on the temperature maps.

The spiral-like shape of the disturbance in the gas becomes apparent when we consider the residuals of Fig.~\ref{002a}. We use the gas densities and temperatures to compute the X-ray emission assuming pure bremsstrahlung ($\propto n_e^2 T^{1/2}$). A $\beta$-model is then fitted to the simulated X-ray map and subtracted from it. The result is shown in Fig.~\ref{002a} with the same arrays of times and inclinations of Fig.~\ref{001a}. We find that the approximate morphology is recovered at $t=1.55$\,Gyr and it seems to be time-sensitive, meaning that the target morphology worsens considerably only a few 100\,Myr earlier or later. Based on a visual inspection, it might be argued that the $i=70\degree$ offers a fair agreement with the observational residuals. Although the other inclinations cannot be confidently ruled out, at low $i$ the north region does not seem ideal, while at $i=90\degree$ the non-axisymmetry seems slightly decreased. Thus we adopt the moment $t=1.55$\,Gyr with $i=70\degree$ as the preferred reference model. Even though it cannot be said to produce strikingly better agreement, it serves as a representative case of a range of acceptable results.

With this in mind, we present a comparison between that simulation model and the observations in Fig.~\ref{005m}. While the simulated temperature map does capture the overall features, it shows a minor shortcoming: the simulated hot gas on the northeast is not as hot as in the observations, although the departure is hardly greater than 0.5--1.0\,keV. Regarding the comparison between residuals in the bottom panels of Fig.~\ref{005m}, again we find a fair overall agreement, with a few drawbacks. In general, the spiral-like shape unfurls in the correct counterclockwise direction -- which necessarily placed the perturbing group on the quadrant where it is found. The dark region of greatest deficit is also reproduced in the correct orientation. The shortcomings are that this simulated dark region is not as intense and not quite as large as in the observations. However this is a sensitive region where the effects of the AGN might be relevant, and that is not included in the present simulation.

Additionally, we present in Fig.~\ref{004a} other gas (and dark matter) properties relating to the best moment of model A, namely: the projected dark matter mass, pressure, entropy ($S = k T n_e^{-2/3}$), temperature, gas density, X-ray emission, fitted $\beta$-model and residuals. It is noticeable that the asymmetry is very mild in all quantities. Temperature maps are often the most sensitive to cold fronts, shocks and general gas disturbances during cluster mergers. Here, the temperature map is the only direct quantity in which the asymmetry is immediately perceptible (apart from the residuals). Yet, it should be noted that the dynamic range of the temperature maps is quite narrow, which contributes to it being more sensitive to small jumps. In cases of a very pronounced sloshing, the outline of the spiral may be hinted at in the density or in the direct X-ray maps. Here, such asymmetry is barely recognizable even a posteriori in those two direct maps. Finally, the residuals from the subtracted $\beta$-model clearly highlight the phenomenon and are one of the most stringent constraints when comparing the simulation to the observation.

\subsubsection{Parameter space}

\begin{figure}
\includegraphics{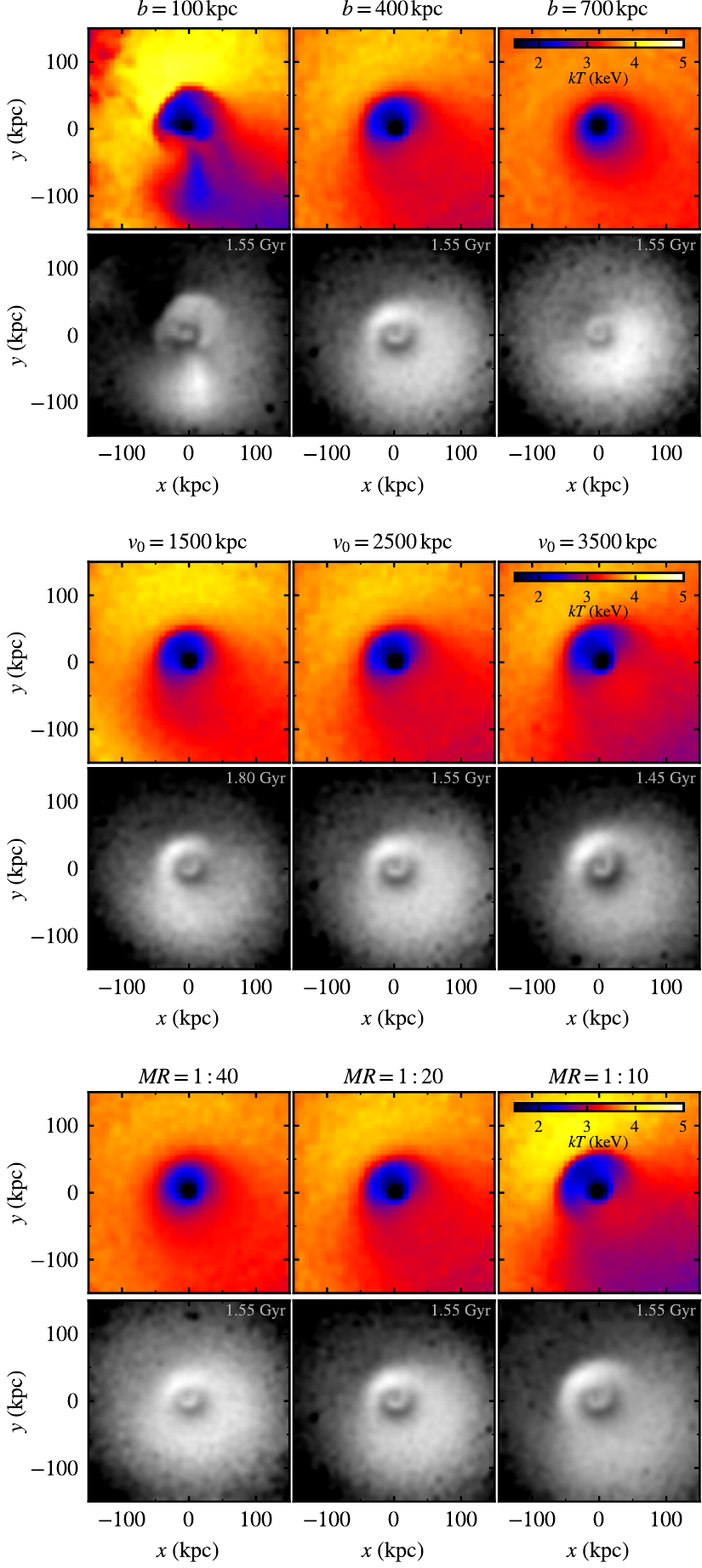}
\caption{Variations around the best model. From top to bottom: impact parameter, initial velocity, and mass ratio, showing temperatures and X-ray residuals for each quantity.}
\label{009m}
\end{figure}

Model A was offered as a representative model within a class of acceptable solutions. Here we present a set of additional model variations. These are meant to serve as a small but systematic exploration of the consequences of varying some of the collision parameters. By keeping all other properties fixed, we highlight the effects of each individual parameter. Table~\ref{tab:variations} presents the model variations, in which we varied: the impact parameter $b$, the relative initial velocity $v_0$, and the mass ratio MR. The results of the additional variations are given in Fig.~\ref{009m}, showing temperature maps and residuals. In each block (from top to bottom: $b$, $v_0$, and MR) the central column corresponds to the default simulation (model A), while the first and third columns correspond to variations around it.

\begin{figure}
\includegraphics{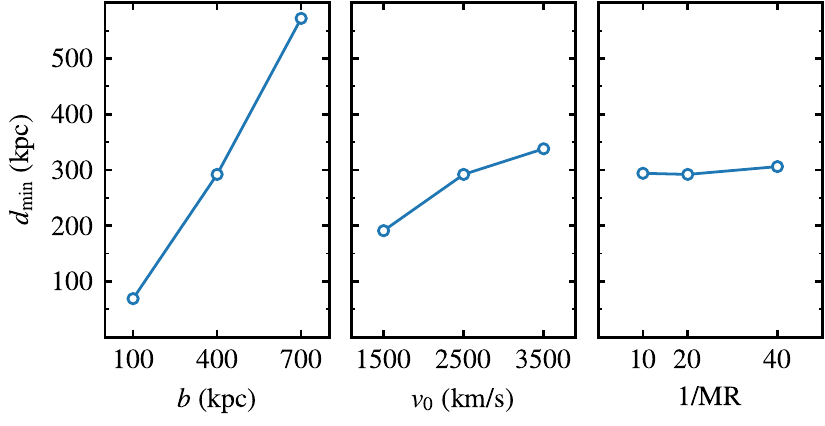}
\caption{Distance between the two clusters at the moment of pericentric passage, for the model variations presented in Table~\ref{tab:variations}.}
\label{000a}
\end{figure}

The different impact parameters are shown in the first block of Fig.~\ref{009m}. Model b100 displays considerably greater disturbance in the gas, because in this case the central passage takes place with a very small minimum separation. Conversely, in model b700, the group passes with a greater separation and thus the perturbation is even milder than in the default model. The values of the minimum separation $d_{\rm min}$, i.e. the distance between the cluster centres at the moment of pericentric passage, are given in Fig.~\ref{000a}. The values of $d_{\rm min}$ are always smaller than $b$. Central passages with $d_{\rm min}$ in the range 100--700\,kpc clearly give rise to noticeably different morphologies. Therefore values towards the edges of this range may be ruled out -- at least for the remaining fixed parameters.

The velocity of the default model was chosen such as to ensure the arrival of the perturber at a distance of $\sim$2.1\,Mpc from the main cluster centre. This velocity is somewhat high. In the cosmological context, major mergers with such velocity would be improbable \citep{Hayashi2006}, but perhaps less so for a low-mass subhalo. In the second block of Fig.~\ref{009m}, the results with different velocities are given. It should be noted that different velocities naturally alter the time scales of the encounter, and thus the best moments of the frames are selected accordingly. Model v1500 provides quite acceptable results. In fact, in the absence of the northeast group constraint, model v1500 might be the preferred model, due to its more plausible velocity. On the other end, model v3500, while physically implausible, is presented as an extreme case of the parameter exploration. The result is that the morphology of the sloshing alone would not be enough to draw conclusions about the velocity. The second panel of Fig.~\ref{000a} shows that the pericentric distance also depends on $v_0$ (bearing in mind that in the $v_0$ variations $b=400$\,kpc is fixed). This is understandable in the sense that a high-speed incoming object will not have its trajectory as easily deflected as a low-speed one. Nevertheless, this effect is relatively modest, since a wide range of velocities (1500--3500\kms) causes $d_{\rm min}$ to vary by only 150\,kpc. As a result, the maps of Fig.~\ref{009m} are not too dissimilar from one another.

Finally, the third block of Fig.~\ref{009m} compares models with different mass ratios. The main cluster is kept fixed and only the mass of the substructure is altered, by a factor of 2 above or below the default model. In the context of our analysis, this is not truly a free parameter, because the mass of the presumed perturber is already constrained by the observations. Nevertheless, this comparison indicates that the mass of the subcluster is an important factor in disturbing the gas. A very low-mass galaxy group would induce an even milder sloshing, while a group twice as massive would accentuate the features. As expected, the pericentric distances in the third panel of Fig.~\ref{000a} are nearly identical for all subscluster masses.

\begin{table}
\caption{Parameters of the initial conditions for the model variations: impact parameter, initial relative velocity and mass ratio.} 
\label{tab:variations}
\begin{center}
\begin{tabular}{cccc}
\hline
model name & $b$ (kpc) & $v_0$ (\kms) & MR   \\
\hline
A          & 400       & 2500         & $1:20$ \\
b100       & 100       & 2500         & $1:20$ \\
b700       & 700       & 2500         & $1:20$ \\
v1500      & 400       & 1500         & $1:20$ \\
v3500      & 400       & 3500         & $1:20$ \\
M40        & 400       & 2500         & $1:40$ \\
M10        & 400       & 2500         & $1:10$ \\
\hline
\end{tabular}
\end{center}
\end{table}

\subsubsection{Dynamics of the dark matter peak}

Here we explore the dynamics of the inner region, focusing on the DM peak. The perturbation caused by the passage of the subcluster affects not only the gas but also the dynamics of the collisionless DM particles in the main cluster core.

Two ways of locating a centroid of the main cluster are: the position of the peak of DM density; and the position of the bottom of the potential well. To decrease numerical noise, we take the median of the 0.01 per cent particles of highest local DM density (or potential). The centroids are both sharply defined, and they coincide on average, within the noise due to numerical resolution. A frame of reference may be defined as the rest frame of the main cluster's centre of mass. At the beginning of the simulation, the origin of this frame of reference coincides with the location of the peaks described above. The external perturbation due to the subcluster introduces a deformation, and the distribution of DM mass in the main cluster is no longer spherically symmetric. Thus, a small offset develops between the density/potential peak and the centre of mass. In a sense, this offset is a measure of the global asymmetry, as the DM halo settles into a non-spherical mass distribution.

\begin{figure}
\includegraphics{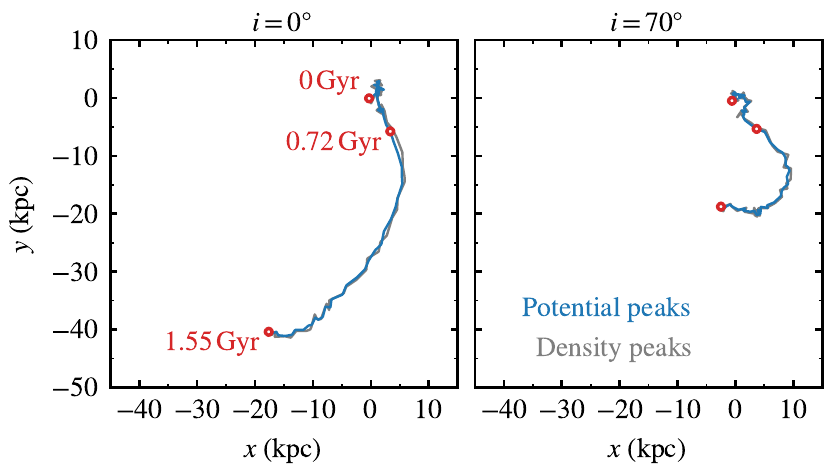}
\caption[]{The inner region of the main cluster, shown in the rest frame defined by the centre of mass of the main cluster. The blue lines correpond to the bottom of the potential well. The gray lines correspond to the locations of the peak of DM density. The red circles mark the beginning of the simulation ($t=0$), the moment of pericentric passage ($t=0.72$\,Gyr) and the best instant of the simulation ($t=1.55$\,Gyr). The left frame is a face-on view and the right frame is seen under an inclination.}
\label{010b}
\end{figure}

The evolution of this offset is shown in Fig.~\ref{010b}. This is analogous to figure 4 from \cite{Ascasibar2006}. At first, before the pericentric passage ($t=0.72$\,Gyr) the density peak is drawn towards the right side of the figure, i.e.~towards the incoming subcluster. The motion of the density peak is later bent towards the left. Thus the density peak would have been displaced -- in the centre-of-mass frame -- by some ${\sim}$40\,kpc in 0.8\,Gyr. Under an inclination of $i=70\degree$, this would translate to about ${\sim}$15\,kpc in the plane of the sky.

\section{Summary and conclusions}
\label{sec:conclusions}

Abell 2199 is a galaxy cluster showing signatures of gas sloshing. In particular, \cite{Nulsen2013} proposed that some features of A2199 might be understood as a sloshing event seen approximately edge-on, pointing out that they are broadly consistent with general sloshing simulations. In this paper, we aimed at producing dedicated simulations tailored to account for the quantitative observed properties of A2199. We wished to evaluate whether the detailed projected morphology of the spiral-like feature was plausible under a large inclination angle.

The Chandra temperature map shows a complex distribution in the very inner part. Although in a larger scale, this cooler region shows some filaments that resembles in morphology the H-$\alpha$ ones seen in HST images. This may be an evidence of the cooling flow towards the central galaxy. The residual map shows a spiral-arm morphology characteristic of sloshing due to a  minor merger event, that was confirmed by the optical analysis and supported by the simulation results. 

From the optical analysis we detect 674 cluster member galaxies sub-divided into a main group and three subgroups, indicating that A2199 is still in the process of virialization. 

We carried out several hydrodynamical $N$-body simulations whose initial conditions were meant to satisfy the observed density profile and temperature profile. We obtained a range of plausible models and offered one to be discussed as a representative case. In that scenario, a galaxy group of virial mass $1.6 \times 10^{13}\Msun$ passes by the main cluster at a mimimum separation of 292\,kpc. This mass is of the same order of magnitude as one of the subgroups identified in the optical analysis. We would be observing the system 0.8\,Gyr after the pericentric passage. The simulation results were projected by an inclination of $i=70\degree$ between the plane of the orbit and the plane of the sky; we found that this inclination is still consistent with the observations.

Since the optical analysis revealed a few substructures with masses of ${\sim}10^{13-14}\Msun$, we investigated whether a simulation model could be found in which the perturber was approximately comparable those substructures. A suitable candidate would be a putative galaxy group located towards the northeast quadrant. Other orientations were difficult to reconcile with the shape of the spiral-like feature in the gas. In any case, the simulated substructure would be currently more than 2\,Mpc away from the cluster core. We argue that the perturber used in the simulations is plausible, because the structures identified by the optical analysis have similar distances from the center. This suggests that the proposed collision is within a realistic regime, although we cannot identify one specific system as the observed counterpart. We were able to find adequate models where several gas properties are recovered within acceptable agreement. However, for the group in question to be considered the responsible for the perturbation, a somewhat high relative initial velocity of 2500\kms was needed in the simulation. If this constraint were lifted -- i.e. if the perturber had the freedom to end at altenative separations --, then models with relative velocity of 1200\kms would account for the morphological features with similar degrees of accuracy.

The age of the collision was estimated as $\sim$0.4\,Gyr by \cite{Nulsen2013}, based on the current extent of the northern plume and on the expected speed of a cold front compared to the local sound speed. Since this relies on a projected plume size with unknown inclination, the age estimate may be regarded as a lower limit. In any case, this is much older than the shocks driven by AGN outbursts. In the model we presented, the age of 0.8\,Gyr is in part a consequence of the requirement that the perturbing group reach a large final separation.

In our reference model, an inclination of $i=70\degree$ was adopted. However, the morphology of the spiral feature was not tightly constrained by the inclination. Even though other inclinations cannot be strongly ruled out, we find that the model is indeed consistent with large inclination angles. Projecting a sloshing spiral under arbitrary inclinations generally gives rise to morphologies that are not trivial to interpret, specially when the line of sight is almost parallel to the orbital plane \citep[e.g.][]{Roediger2011}. In this sense, it is not obvious a priori what the precise sloshing signature would be for a given cluster model under large inclination. The fact that this particular model, constructed to match the observations, gives rise to a comparable spiral feature indicates the consistency of the scenario.

The sloshing signature is generally characterized by cool gas that was driven out of the cluster core. In the temperature maps, this tends to produce a non-axisymmetric feature. In our specific model, the disturbance is somewhat mild, due to the low mass of the perturbing group. Coupled with the inclination, this resulted in a temperature map with a slightly asymmetric cool core and a higher temperature towards the northeast. These properties, as well as the temperature ranges, are globally in agreement with the observations, namely roughly 2\,keV in the cool core and 4\,keV in the northeast.

Yet the sloshing spiral is more noticeable in the maps of residuals, i.e. the fitted $\beta$ models subtracted from the simulated X-ray emission. The simulated spiral is slightly smaller in extent than the observed one. Although the agreement is not perfect, the overall shape and orientation of the excess are adequate, as well as the dark region of deficit. Disagreements in the scale of a few 10\,kpc are expected, since this simulation does not include the effects of the AGN, and thus it could not have been expected to reproduce small-scale features caused by the injection of energy from the radio jets. Nevertheless, the fact that such an idealized model spontaneously gave rise such to such a detailed feature is a good indication of the general consistency of the scenario and the mechanism.

The exploration of the parameter space around the default model illustrates the role played by each parameter and also suggests ranges of confidence. All other quantities having been fixed, the impact parameter $b$ is naturally related to $d_{\rm min}$, the minimum separation at the instant of pericentric passage. We found that models with $d_{\rm min}\sim300$\,kpc gave the best results, while $d_{\rm min}\sim100$\,kpc or $d_{\rm min}\sim600$\,kpc would cause, respectively, too violent or too mild disturbances. The initial velocities also affect the pericentric separation, but to a lesser degree. The high velocity (2500\kms) of the default model was imposed by the constraint of matching the perturber position, but smaller velocities (1200\kms) would produce acceptable results otherwise. The mass ratio of 1:20 was also adopted because of the assumption of the northeast group as the perturber. A substructure more massive by a factor of 2 would also be adequate.

Finally, we measured the offset of the dark matter peak of the main cluster, with respect to its centre-of-mass resframe. Due to the gravitational perturbation, the peak was displaced by roughly ${\sim}$40\,kpc in the 0.8\,Gyr since pericentric passage. Seen in projection, this would mean an offset of ${\sim}$15\,kpc in the plane of the sky.

Some simplifications are present in such idealized simulations: the initial clusters are perfectly spherical, there are no substructures, and the phenomenon is modeled based on one single encounter. Such assumptions are needed to isolate the phenomenon under study. However, real clusters naturally have a much more complicated history, undergoing multiple merging events. In this sense, it cannot be decisively argued that the simulated group is the sole cause of the currently observed sloshing. Rather, we conclude that it is a plausible candidate, since a consistent orbit can be found; but noting that other encounters with objects of similar mass could also have given rise to the sloshing features.

Finally, the off-axis collision between A2199 and a galaxy group explored in this work could also induce gravitational perturbations in the galaxies as well, modifying their original orbits around A2199's gravitational centre. Indeed, \citet{1983ApJ...271..575B} argued in favour of a possible motion of 3C\,338 as the main driver of the unusual and complex structures seen at kiloparsec-scale images of this source. We will analyse this scenario for 3C\,338 in the context of three-dimensional hydrodynamic simulations in a forthcoming paper.

\section*{Acknowledgements}

The authors acknowledge the National Laboratory for Scientific Computing (LNCC/MCTI, Brazil) for providing HPC resources of the SDumont supercomputer, which have contributed to the research results reported within this paper. REGM acknowledges support from the Brazilian agency \textit {Conselho Nacional de Desenvolvimento Cient\'ifico e Tecnol\'ogico} (CNPq) through grants 303426/2018-7, 406908/2018-4 and 307205/2021-5. REGM acknowledges support from \textit{Funda\c c\~ao de Apoio \`a Ci\^encia, Tecnologia e Inova\c c\~ao do Paran\'a} through grant 18.148.096-3 -- NAPI \textit{Fen\^omenos Extremos do Universo}. ASRA thanks Coordena\c{c}\~ao de Aperfei\c{c}oamento de Pessoal de N\'\i vel Superior (CAPES) for financial support.

\section*{Data availability}
The data underlying this article will be shared on reasonable request to the corresponding author.

\bibliographystyle{mnras.bst}
\bibliography{A2199.bib}

\begin{thebibliography}{}
\makeatletter
\relax
\def\mn@urlcharsother{\let\do\@makeother \do\$\do\&\do\#\do\^\do\_\do\%\do\~}
\def\mn@doi{\begingroup\mn@urlcharsother \@ifnextchar [ {\mn@doi@}
  {\mn@doi@[]}}
\def\mn@doi@[#1]#2{\def\@tempa{#1}\ifx\@tempa\@empty \href
  {http://dx.doi.org/#2} {doi:#2}\else \href {http://dx.doi.org/#2} {#1}\fi
  \endgroup}
\def\mn@eprint#1#2{\mn@eprint@#1:#2::\@nil}
\def\mn@eprint@arXiv#1{\href {http://arxiv.org/abs/#1} {{\tt arXiv:#1}}}
\def\mn@eprint@dblp#1{\href {http://dblp.uni-trier.de/rec/bibtex/#1.xml}
  {dblp:#1}}
\def\mn@eprint@#1:#2:#3:#4\@nil{\def\@tempa {#1}\def\@tempb {#2}\def\@tempc
  {#3}\ifx \@tempc \@empty \let \@tempc \@tempb \let \@tempb \@tempa \fi \ifx
  \@tempb \@empty \def\@tempb {arXiv}\fi \@ifundefined
  {mn@eprint@\@tempb}{\@tempb:\@tempc}{\expandafter \expandafter \csname
  mn@eprint@\@tempb\endcsname \expandafter{\@tempc}}}

\bibitem[\protect\citeauthoryear{{Alden}, {Hallman}, {Rapetti}, {Burns}  \&
  {Datta}}{{Alden} et~al.}{2019}]{Clusteryxt19}
{Alden} B.,  {Hallman} E.~J.,  {Rapetti} D.,  {Burns} J.~O.,   {Datta} A.,
  2019, \mn@doi [Astronomy and Computing] {10.1016/j.ascom.2019.04.001}, \href
  {https://ui.adsabs.harvard.edu/abs/2019A&C....27..147A} {27, 147}

\bibitem[\protect\citeauthoryear{{Allen}, {Dunn}, {Fabian}, {Taylor}  \&
  {Reynolds}}{{Allen} et~al.}{2006}]{2006MNRAS.372...21A}
{Allen} S.~W.,  {Dunn} R.~J.~H.,  {Fabian} A.~C.,  {Taylor} G.~B.,   {Reynolds}
  C.~S.,  2006, \mn@doi [\mnras] {10.1111/j.1365-2966.2006.10778.x}, \href
  {https://ui.adsabs.harvard.edu/abs/2006MNRAS.372...21A} {372, 21}

\bibitem[\protect\citeauthoryear{{Ascasibar} \& {Markevitch}}{{Ascasibar} \&
  {Markevitch}}{2006a}]{AscasibarMarkevitch2006}
{Ascasibar} Y.,  {Markevitch} M.,  2006a, \mn@doi [\apj] {10.1086/506508},
  \href {http://adsabs.harvard.edu/abs/2006ApJ...650..102A} {650, 102}

\bibitem[\protect\citeauthoryear{{Ascasibar} \& {Markevitch}}{{Ascasibar} \&
  {Markevitch}}{2006b}]{Ascasibar2006}
{Ascasibar} Y.,  {Markevitch} M.,  2006b, \mn@doi [\apj] {10.1086/506508},
  \href {https://ui.adsabs.harvard.edu/abs/2006ApJ...650..102A} {650, 102}

\bibitem[\protect\citeauthoryear{{Asplund}, {Grevesse}, {Sauval}  \&
  {Scott}}{{Asplund} et~al.}{2009}]{Asplund09}
{Asplund} M.,  {Grevesse} N.,  {Sauval} A.~J.,   {Scott} P.,  2009, \mn@doi
  [\araa] {10.1146/annurev.astro.46.060407.145222}, \href
  {https://ui.adsabs.harvard.edu/abs/2009ARA&A..47..481A} {47, 481}

\bibitem[\protect\citeauthoryear{{B{\^\i}rzan}, {Rafferty}, {McNamara}, {Wise}
  \& {Nulsen}}{{B{\^\i}rzan} et~al.}{2004}]{2004ApJ...607..800B}
{B{\^\i}rzan} L.,  {Rafferty} D.~A.,  {McNamara} B.~R.,  {Wise} M.~W.,
  {Nulsen} P.~E.~J.,  2004, \mn@doi [\apj] {10.1086/383519}, \href
  {https://ui.adsabs.harvard.edu/abs/2004ApJ...607..800B} {607, 800}

\bibitem[\protect\citeauthoryear{{Blanton}, {Randall}, {Douglass}, {Sarazin},
  {Clarke}  \& {McNamara}}{{Blanton} et~al.}{2009}]{Blanton2009}
{Blanton} E.~L.,  {Randall} S.~W.,  {Douglass} E.~M.,  {Sarazin} C.~L.,
  {Clarke} T.~E.,   {McNamara} B.~R.,  2009, \mn@doi [\apjl]
  {10.1088/0004-637X/697/2/L95}, \href
  {http://adsabs.harvard.edu/abs/2009ApJ...697L..95B} {697, L95}

\bibitem[\protect\citeauthoryear{{Blanton}, {Randall}, {Clarke}, {Sarazin},
  {McNamara}, {Douglass}  \& {McDonald}}{{Blanton} et~al.}{2011}]{Blanton2011}
{Blanton} E.~L.,  {Randall} S.~W.,  {Clarke} T.~E.,  {Sarazin} C.~L.,
  {McNamara} B.~R.,  {Douglass} E.~M.,   {McDonald} M.,  2011, \mn@doi [\apj]
  {10.1088/0004-637X/737/2/99}, \href
  {http://adsabs.harvard.edu/abs/2011ApJ...737...99B} {737, 99}

\bibitem[\protect\citeauthoryear{{Burns}, {Schwendeman}  \& {White}}{{Burns}
  et~al.}{1983}]{1983ApJ...271..575B}
{Burns} J.~O.,  {Schwendeman} E.,   {White} R.~A.,  1983, \mn@doi [\apj]
  {10.1086/161224}, \href
  {https://ui.adsabs.harvard.edu/abs/1983ApJ...271..575B} {271, 575}

\bibitem[\protect\citeauthoryear{{Chadayammuri}, {ZuHone}, {Nulsen}, {Nagai},
  {Felix}, {Andrade-Santos}, {King}  \& {Russell}}{{Chadayammuri}
  et~al.}{2021}]{Chadayammuri2021}
{Chadayammuri} U.,  {ZuHone} J.,  {Nulsen} P.,  {Nagai} D.,  {Felix} S.,
  {Andrade-Santos} F.,  {King} L.,   {Russell} H.,  2021, \mn@doi [\mnras]
  {10.1093/mnras/stab2629}, \href
  {https://ui.adsabs.harvard.edu/abs/2021MNRAS.tmp.2872C} {}

\bibitem[\protect\citeauthoryear{{Churazov}, {Forman}, {Jones}  \&
  {B{\"o}hringer}}{{Churazov} et~al.}{2003}]{Churazov2003}
{Churazov} E.,  {Forman} W.,  {Jones} C.,   {B{\"o}hringer} H.,  2003, \mn@doi
  [\apj] {10.1086/374923}, \href
  {http://adsabs.harvard.edu/abs/2003ApJ...590..225C} {590, 225}

\bibitem[\protect\citeauthoryear{{Clarke}, {Blanton}  \& {Sarazin}}{{Clarke}
  et~al.}{2004}]{Clarke2004}
{Clarke} T.~E.,  {Blanton} E.~L.,   {Sarazin} C.~L.,  2004, \mn@doi [\apj]
  {10.1086/424911}, \href {http://adsabs.harvard.edu/abs/2004ApJ...616..178C}
  {616, 178}

\bibitem[\protect\citeauthoryear{{Dehnen}}{{Dehnen}}{1993}]{Dehnen1993}
{Dehnen} W.,  1993, \mnras, \href
  {http://adsabs.harvard.edu/abs/1993MNRAS.265..250D} {265, 250}

\bibitem[\protect\citeauthoryear{{Diaferio}}{{Diaferio}}{1999}]{Diaferio99}
{Diaferio} A.,  1999, \mn@doi [\mnras] {10.1046/j.1365-8711.1999.02864.x},
  \href {https://ui.adsabs.harvard.edu/abs/1999MNRAS.309..610D} {309, 610}

\bibitem[\protect\citeauthoryear{{Diaferio} \& {Geller}}{{Diaferio} \&
  {Geller}}{1997}]{Diaferio97}
{Diaferio} A.,  {Geller} M.~J.,  1997, \mn@doi [\apj] {10.1086/304075}, \href
  {https://ui.adsabs.harvard.edu/abs/1997ApJ...481..633D} {481, 633}

\bibitem[\protect\citeauthoryear{{Doubrawa}, {Machado}, {Lagan{\'a}}, {Lima
  Neto}, {Monteiro-Oliveira}  \& {Cypriano}}{{Doubrawa}
  et~al.}{2020}]{Doubrawa2020}
{Doubrawa} L.,  {Machado} R.~E.~G.,  {Lagan{\'a}} T.~F.,  {Lima Neto} G.~B.,
  {Monteiro-Oliveira} R.,   {Cypriano} E.~S.,  2020, \mn@doi [\mnras]
  {10.1093/mnras/staa1051}, \href
  {https://ui.adsabs.harvard.edu/abs/2020MNRAS.495.2022D} {495, 2022}

\bibitem[\protect\citeauthoryear{{Douglass}, {Blanton}, {Randall}, {Clarke},
  {Edwards}, {Sabry}  \& {ZuHone}}{{Douglass} et~al.}{2018}]{Douglass2018}
{Douglass} E.~M.,  {Blanton} E.~L.,  {Randall} S.~W.,  {Clarke} T.~E.,
  {Edwards} L.~O.~V.,  {Sabry} Z.,   {ZuHone} J.~A.,  2018, \mn@doi [\apj]
  {10.3847/1538-4357/aae9e7}, \href
  {https://ui.adsabs.harvard.edu/abs/2018ApJ...868..121D} {868, 121}

\bibitem[\protect\citeauthoryear{{Duffy}, {Schaye}, {Kay}  \& {Dalla
  Vecchia}}{{Duffy} et~al.}{2008}]{Duffy2008}
{Duffy} A.~R.,  {Schaye} J.,  {Kay} S.~T.,   {Dalla Vecchia} C.,  2008, \mn@doi
  [\mnras] {10.1111/j.1745-3933.2008.00537.x}, \href
  {https://ui.adsabs.harvard.edu/abs/2008MNRAS.390L..64D} {390, L64}

\bibitem[\protect\citeauthoryear{{Dupke}, {Mirabal}, {Bregman}  \&
  {Evrard}}{{Dupke} et~al.}{2007}]{Dupke2007}
{Dupke} R.~A.,  {Mirabal} N.,  {Bregman} J.~N.,   {Evrard} A.~E.,  2007,
  \mn@doi [\apj] {10.1086/520708}, \href
  {https://ui.adsabs.harvard.edu/abs/2007ApJ...668..781D} {668, 781}

\bibitem[\protect\citeauthoryear{{Durret}, {Lagan{\'a}}, {Adami}  \&
  {Bertin}}{{Durret} et~al.}{2010}]{Durret10}
{Durret} F.,  {Lagan{\'a}} T.~F.,  {Adami} C.,   {Bertin} E.,  2010, \mn@doi
  [\aap] {10.1051/0004-6361/201014566}, \href
  {https://ui.adsabs.harvard.edu/abs/2010A&A...517A..94D} {517, A94}

\bibitem[\protect\citeauthoryear{{Durret}, {Lagan{\'a}}  \& {Haider}}{{Durret}
  et~al.}{2011}]{Durret11}
{Durret} F.,  {Lagan{\'a}} T.~F.,   {Haider} M.,  2011, \mn@doi [\aap]
  {10.1051/0004-6361/201015978}, \href
  {https://ui.adsabs.harvard.edu/abs/2011A&A...529A..38D} {529, A38}

\bibitem[\protect\citeauthoryear{Fraley \& Raftery}{Fraley \&
  Raftery}{2006}]{Fraley06}
Fraley C.,  Raftery A.~E.,  2006, Technical Report~504, MCLUST Version 3 for R:
  Normal Mixture Modeling and Model-Based Clustering.
University of Washington, Department of Statistics

\bibitem[\protect\citeauthoryear{{Gastaldello}, {Di Gesu}, {Ghizzardi}
  et~al.}{{Gastaldello} et~al.}{2013}]{Gastaldello2013}
{Gastaldello} F.,  {Di Gesu} L.,  {Ghizzardi} S.,   et~al., 2013, \mn@doi
  [\apj] {10.1088/0004-637X/770/1/56}, \href
  {http://adsabs.harvard.edu/abs/2013ApJ...770...56G} {770, 56}

\bibitem[\protect\citeauthoryear{{Ghizzardi}, {Rossetti}  \&
  {Molendi}}{{Ghizzardi} et~al.}{2010}]{Ghizzardi2010}
{Ghizzardi} S.,  {Rossetti} M.,   {Molendi} S.,  2010, \mn@doi [\aap]
  {10.1051/0004-6361/200912496}, \href
  {http://adsabs.harvard.edu/abs/2010A%26A...516A..32G} {516, A32}

\bibitem[\protect\citeauthoryear{{Ghizzardi}, {De Grandi}  \&
  {Molendi}}{{Ghizzardi} et~al.}{2014}]{Ghizzardi2014}
{Ghizzardi} S.,  {De Grandi} S.,   {Molendi} S.,  2014, \mn@doi [\aap]
  {10.1051/0004-6361/201424016}, \href
  {https://ui.adsabs.harvard.edu/abs/2014A&A...570A.117G} {570, A117}

\bibitem[\protect\citeauthoryear{Gifford \& Miller}{Gifford \&
  Miller}{2013}]{Gifford-Miller13}
Gifford D.,  Miller C.~J.,  2013, \mn@doi [The Astrophysical Journal]
  {10.1088/2041-8205/768/2/l32}, 768, L32

\bibitem[\protect\citeauthoryear{Gifford, Miller  \& Kern}{Gifford
  et~al.}{2013}]{Gifford13}
Gifford D.,  Miller C.,   Kern N.,  2013, \mn@doi [The Astrophysical Journal]
  {10.1088/0004-637x/773/2/116}, 773, 116

\bibitem[\protect\citeauthoryear{{Hayashi} \& {White}}{{Hayashi} \&
  {White}}{2006}]{Hayashi2006}
{Hayashi} E.,  {White} S. D.~M.,  2006, \mn@doi [\mnras]
  {10.1111/j.1745-3933.2006.00184.x}, \href
  {https://ui.adsabs.harvard.edu/abs/2006MNRAS.370L..38H} {370, L38}

\bibitem[\protect\citeauthoryear{{Hernquist}}{{Hernquist}}{1990}]{Hernquist1990}
{Hernquist} L.,  1990, \mn@doi [\apj] {10.1086/168845}, \href
  {http://adsabs.harvard.edu/abs/1990ApJ...356..359H} {356, 359}

\bibitem[\protect\citeauthoryear{{Ichinohe}, {Simionescu}, {Werner}, {Fabian}
  \& {Takahashi}}{{Ichinohe} et~al.}{2019}]{Ichinohe2019}
{Ichinohe} Y.,  {Simionescu} A.,  {Werner} N.,  {Fabian} A.~C.,   {Takahashi}
  T.,  2019, \mn@doi [\mnras] {10.1093/mnras/sty3257}, \href
  {https://ui.adsabs.harvard.edu/abs/2019MNRAS.483.1744I} {483, 1744}

\bibitem[\protect\citeauthoryear{{Johnson}, {Markevitch}, {Wegner}, {Jones}  \&
  {Forman}}{{Johnson} et~al.}{2010}]{Johnson2010}
{Johnson} R.~E.,  {Markevitch} M.,  {Wegner} G.~A.,  {Jones} C.,   {Forman}
  W.~R.,  2010, \mn@doi [\apj] {10.1088/0004-637X/710/2/1776}, \href
  {http://adsabs.harvard.edu/abs/2010ApJ...710.1776J} {710, 1776}

\bibitem[\protect\citeauthoryear{{Johnson}, {Zuhone}, {Jones}, {Forman}  \&
  {Markevitch}}{{Johnson} et~al.}{2012}]{Johnson2012}
{Johnson} R.~E.,  {Zuhone} J.,  {Jones} C.,  {Forman} W.~R.,   {Markevitch} M.,
   2012, \mn@doi [\apj] {10.1088/0004-637X/751/2/95}, \href
  {http://adsabs.harvard.edu/abs/2012ApJ...751...95J} {751, 95}

\bibitem[\protect\citeauthoryear{{Johnstone}, {Allen}, {Fabian}  \&
  {Sanders}}{{Johnstone} et~al.}{2002}]{2002MNRAS.336..299J}
{Johnstone} R.~M.,  {Allen} S.~W.,  {Fabian} A.~C.,   {Sanders} J.~S.,  2002,
  \mn@doi [\mnras] {10.1046/j.1365-8711.2002.05743.x}, \href
  {https://ui.adsabs.harvard.edu/abs/2002MNRAS.336..299J} {336, 299}

\bibitem[\protect\citeauthoryear{Kass \& Raftery}{Kass \&
  Raftery}{1995}]{KassRaftery1995}
Kass R.~E.,  Raftery A.~E.,  1995, Journal of the American Statistical
  Association, 90, 773

\bibitem[\protect\citeauthoryear{{Kazantzidis}, {Kravtsov}, {Zentner},
  {Allgood}, {Nagai}  \& {Moore}}{{Kazantzidis} et~al.}{2004}]{Kazantzidis2004}
{Kazantzidis} S.,  {Kravtsov} A.~V.,  {Zentner} A.~R.,  {Allgood} B.,  {Nagai}
  D.,   {Moore} B.,  2004, \mn@doi [\apjl] {10.1086/423992}, \href
  {http://adsabs.harvard.edu/abs/2004ApJ...611L..73K} {611, L73}

\bibitem[\protect\citeauthoryear{{Kazantzidis}, {Zentner}  \&
  {Kravtsov}}{{Kazantzidis} et~al.}{2006}]{Kazantzidis2006}
{Kazantzidis} S.,  {Zentner} A.~R.,   {Kravtsov} A.~V.,  2006, \mn@doi [\apj]
  {10.1086/500579}, \href {http://adsabs.harvard.edu/abs/2006ApJ...641..647K}
  {641, 647}

\bibitem[\protect\citeauthoryear{{Lagan{\'a}}, {Lima Neto}, {Andrade-Santos}
  \& {Cypriano}}{{Lagan{\'a}} et~al.}{2008}]{Lagana08}
{Lagan{\'a}} T.~F.,  {Lima Neto} G.~B.,  {Andrade-Santos} F.,   {Cypriano}
  E.~S.,  2008, \mn@doi [\aap] {10.1051/0004-6361:20079168}, \href
  {https://ui.adsabs.harvard.edu/abs/2008A&A...485..633L} {485, 633}

\bibitem[\protect\citeauthoryear{{Lagan{\'a}}, {Andrade-Santos}  \& {Lima
  Neto}}{{Lagan{\'a}} et~al.}{2010}]{Lagana2010}
{Lagan{\'a}} T.~F.,  {Andrade-Santos} F.,   {Lima Neto} G.~B.,  2010, \mn@doi
  [\aap] {10.1051/0004-6361/200913180}, \href
  {http://adsabs.harvard.edu/abs/2010A%26A...511A..15L} {511, A15}

\bibitem[\protect\citeauthoryear{{Lagan{\'a}}, {Martinet}, {Durret}, {Lima
  Neto}, {Maughan}  \& {Zhang}}{{Lagan{\'a}} et~al.}{2013}]{Lagana13}
{Lagan{\'a}} T.~F.,  {Martinet} N.,  {Durret} F.,  {Lima Neto} G.~B.,
  {Maughan} B.,   {Zhang} Y.-Y.,  2013, \mn@doi [\aap]
  {10.1051/0004-6361/201220423}, \href
  {http://adsabs.harvard.edu/abs/2013A%26A...555A..66L} {555, A66}

\bibitem[\protect\citeauthoryear{{Lagan{\'a}}, {Durret}  \&
  {Lopes}}{{Lagan{\'a}} et~al.}{2019a}]{Lagana19}
{Lagan{\'a}} T.~F.,  {Durret} F.,   {Lopes} P.~A.~A.,  2019a, \mn@doi [\mnras]
  {10.1093/mnras/stz148}, \href
  {https://ui.adsabs.harvard.edu/abs/2019MNRAS.484.2807L} {484, 2807}

\bibitem[\protect\citeauthoryear{{Lagan{\'a}}, {Souza}, {Machado}, {Volert}  \&
  {Lopes}}{{Lagan{\'a}} et~al.}{2019b}]{Lagana2019}
{Lagan{\'a}} T.~F.,  {Souza} G.~S.,  {Machado} R.~E.~G.,  {Volert} R.~C.,
  {Lopes} P.~A.~A.,  2019b, \mn@doi [\mnras] {10.1093/mnras/stz1575}, \href
  {https://ui.adsabs.harvard.edu/abs/2019MNRAS.487.3922L} {487, 3922}

\bibitem[\protect\citeauthoryear{{Lal} et~al.,}{{Lal} et~al.}{2013}]{Lal2013}
{Lal} D.~V.,  et~al., 2013, \mn@doi [\apj] {10.1088/0004-637X/764/1/83}, \href
  {http://adsabs.harvard.edu/abs/2013ApJ...764...83L} {764, 83}

\bibitem[\protect\citeauthoryear{{Lee}, {Hwang}, {Lee}, {Ko}, {Sohn}, {Shim}
  \& {Diaferio}}{{Lee} et~al.}{2015}]{Lee2015}
{Lee} G.-H.,  {Hwang} H.~S.,  {Lee} M.~G.,  {Ko} J.,  {Sohn} J.,  {Shim} H.,
  {Diaferio} A.,  2015, \mn@doi [\apj] {10.1088/0004-637X/800/2/80}, \href
  {https://ui.adsabs.harvard.edu/abs/2015ApJ...800...80L} {800, 80}

\bibitem[\protect\citeauthoryear{{Liu}, {Yu}, {Diaferio}, {Tozzi}, {Hwang},
  {Umetsu}, {Okabe}  \& {Yang}}{{Liu} et~al.}{2018}]{Liu2018}
{Liu} A.,  {Yu} H.,  {Diaferio} A.,  {Tozzi} P.,  {Hwang} H.~S.,  {Umetsu} K.,
  {Okabe} N.,   {Yang} L.-L.,  2018, \mn@doi [\apj] {10.3847/1538-4357/aad090},
  \href {https://ui.adsabs.harvard.edu/abs/2018ApJ...863..102L} {863, 102}

\bibitem[\protect\citeauthoryear{{Louren{\c{c}}o} et~al.,}{{Louren{\c{c}}o}
  et~al.}{2020}]{Ana20}
{Louren{\c{c}}o} A. C.~C.,  et~al., 2020, \mn@doi [\mnras]
  {10.1093/mnras/staa2464}, \href
  {https://ui.adsabs.harvard.edu/abs/2020MNRAS.498..835L} {498, 835}

\bibitem[\protect\citeauthoryear{{Lovisari}, {Reiprich}  \&
  {Schellenberger}}{{Lovisari} et~al.}{2015}]{Lovisari2015}
{Lovisari} L.,  {Reiprich} T.~H.,   {Schellenberger} G.,  2015, \mn@doi [\aap]
  {10.1051/0004-6361/201423954}, \href
  {https://ui.adsabs.harvard.edu/abs/2015A&A...573A.118L} {573, A118}

\bibitem[\protect\citeauthoryear{{Machado} \& {Lima Neto}}{{Machado} \& {Lima
  Neto}}{2013}]{Machado2013}
{Machado} R.~E.~G.,  {Lima Neto} G.~B.,  2013, \mn@doi [\mnras]
  {10.1093/mnras/stt127}, \href
  {http://adsabs.harvard.edu/abs/2013MNRAS.430.3249M} {430, 3249}

\bibitem[\protect\citeauthoryear{{Machado} \& {Lima Neto}}{{Machado} \& {Lima
  Neto}}{2015}]{Machado2015}
{Machado} R.~E.~G.,  {Lima Neto} G.~B.,  2015, \mn@doi [\mnras]
  {10.1093/mnras/stu2669}, \href
  {http://adsabs.harvard.edu/abs/2015MNRAS.447.2915M} {447, 2915}

\bibitem[\protect\citeauthoryear{{Markevitch} \& {Vikhlinin}}{{Markevitch} \&
  {Vikhlinin}}{2007}]{Markevitch2007}
{Markevitch} M.,  {Vikhlinin} A.,  2007, \mn@doi [\physrep]
  {10.1016/j.physrep.2007.01.001}, \href
  {http://adsabs.harvard.edu/abs/2007PhR...443....1M} {443, 1}

\bibitem[\protect\citeauthoryear{{Markevitch}, {Vikhlinin}  \&
  {Mazzotta}}{{Markevitch} et~al.}{2001}]{Markevitch2001}
{Markevitch} M.,  {Vikhlinin} A.,   {Mazzotta} P.,  2001, \mn@doi [\apjl]
  {10.1086/337973}, \href {http://adsabs.harvard.edu/abs/2001ApJ...562L.153M}
  {562, L153}

\bibitem[\protect\citeauthoryear{{Mirakhor} \& {Walker}}{{Mirakhor} \&
  {Walker}}{2020}]{Mirakhor2020}
{Mirakhor} M.~S.,  {Walker} S.~A.,  2020, \mn@doi [\mnras]
  {10.1093/mnras/staa2204}, \href
  {https://ui.adsabs.harvard.edu/abs/2020MNRAS.497.3943M} {497, 3943}

\bibitem[\protect\citeauthoryear{{Monteiro-Oliveira}, {Doubrawa}, {Machado},
  {Lima Neto}, {Castejon}  \& {Cypriano}}{{Monteiro-Oliveira}
  et~al.}{2020}]{MonteiroOliveira2020}
{Monteiro-Oliveira} R.,  {Doubrawa} L.,  {Machado} R.~E.~G.,  {Lima Neto}
  G.~B.,  {Castejon} M.,   {Cypriano} E.~S.,  2020, \mn@doi [\mnras]
  {10.1093/mnras/staa1218}, \href
  {https://ui.adsabs.harvard.edu/abs/2020MNRAS.495.2007M} {495, 2007}

\bibitem[\protect\citeauthoryear{{Nulsen} et~al.,}{{Nulsen}
  et~al.}{2013}]{Nulsen2013}
{Nulsen} P. E.~J.,  et~al., 2013, \mn@doi [\apj] {10.1088/0004-637X/775/2/117},
  \href {https://ui.adsabs.harvard.edu/abs/2013ApJ...775..117N} {775, 117}

\bibitem[\protect\citeauthoryear{{Oegerle} \& {Hill}}{{Oegerle} \&
  {Hill}}{2001}]{2001AJ....122.2858O}
{Oegerle} W.~R.,  {Hill} J.~M.,  2001, \mn@doi [\aj] {10.1086/323536}, \href
  {https://ui.adsabs.harvard.edu/abs/2001AJ....122.2858O} {122, 2858}

\bibitem[\protect\citeauthoryear{{Owen} \& {Eilek}}{{Owen} \&
  {Eilek}}{1998}]{1998ApJ...493...73O}
{Owen} F.~N.,  {Eilek} J.~A.,  1998, \mn@doi [\apj] {10.1086/305092}, \href
  {https://ui.adsabs.harvard.edu/abs/1998ApJ...493...73O} {493, 73}

\bibitem[\protect\citeauthoryear{{Owers}, {Nulsen}, {Couch}  \&
  {Markevitch}}{{Owers} et~al.}{2009}]{Owers2009}
{Owers} M.~S.,  {Nulsen} P.~E.~J.,  {Couch} W.~J.,   {Markevitch} M.,  2009,
  \mn@doi [\apj] {10.1088/0004-637X/704/2/1349}, \href
  {http://adsabs.harvard.edu/abs/2009ApJ...704.1349O} {704, 1349}

\bibitem[\protect\citeauthoryear{{Owers}, {Nulsen}  \& {Couch}}{{Owers}
  et~al.}{2011}]{Owers2011}
{Owers} M.~S.,  {Nulsen} P. E.~J.,   {Couch} W.~J.,  2011, \mn@doi [\apj]
  {10.1088/0004-637X/741/2/122}, \href
  {https://ui.adsabs.harvard.edu/abs/2011ApJ...741..122O} {741, 122}

\bibitem[\protect\citeauthoryear{{Paterno-Mahler}, {Blanton}, {Randall}  \&
  {Clarke}}{{Paterno-Mahler} et~al.}{2013}]{PaternoMahler2013}
{Paterno-Mahler} R.,  {Blanton} E.~L.,  {Randall} S.~W.,   {Clarke} T.~E.,
  2013, \mn@doi [\apj] {10.1088/0004-637X/773/2/114}, \href
  {http://adsabs.harvard.edu/abs/2013ApJ...773..114P} {773, 114}

\bibitem[\protect\citeauthoryear{{Peres}, {Fabian}, {Edge}, {Allen},
  {Johnstone}  \& {White}}{{Peres} et~al.}{1998}]{1998MNRAS.298..416P}
{Peres} C.~B.,  {Fabian} A.~C.,  {Edge} A.~C.,  {Allen} S.~W.,  {Johnstone}
  R.~M.,   {White} D.~A.,  1998, \mn@doi [\mnras]
  {10.1046/j.1365-8711.1998.01624.x}, \href
  {https://ui.adsabs.harvard.edu/abs/1998MNRAS.298..416P} {298, 416}

\bibitem[\protect\citeauthoryear{{Piffaretti}, {Arnaud}, {Pratt},
  {Pointecouteau}  \& {Melin}}{{Piffaretti} et~al.}{2011a}]{Piffaretti11}
{Piffaretti} R.,  {Arnaud} M.,  {Pratt} G.~W.,  {Pointecouteau} E.,   {Melin}
  J.~B.,  2011a, \mn@doi [\aap] {10.1051/0004-6361/201015377}, \href
  {https://ui.adsabs.harvard.edu/abs/2011A&A...534A.109P} {534, A109}

\bibitem[\protect\citeauthoryear{{Piffaretti}, {Arnaud}, {Pratt},
  {Pointecouteau}  \& {Melin}}{{Piffaretti} et~al.}{2011b}]{Piffaretti2011}
{Piffaretti} R.,  {Arnaud} M.,  {Pratt} G.~W.,  {Pointecouteau} E.,   {Melin}
  J.~B.,  2011b, \mn@doi [\aap] {10.1051/0004-6361/201015377}, \href
  {https://ui.adsabs.harvard.edu/abs/2011A&A...534A.109P} {534, A109}

\bibitem[\protect\citeauthoryear{{Randall}, {Jones}, {Markevitch}, {Blanton},
  {Nulsen}  \& {Forman}}{{Randall} et~al.}{2009}]{Randall2009}
{Randall} S.~W.,  {Jones} C.,  {Markevitch} M.,  {Blanton} E.~L.,  {Nulsen}
  P.~E.~J.,   {Forman} W.~R.,  2009, \mn@doi [\apj]
  {10.1088/0004-637X/700/2/1404}, \href
  {http://adsabs.harvard.edu/abs/2009ApJ...700.1404R} {700, 1404}

\bibitem[\protect\citeauthoryear{{Read} \& {Ponman}}{{Read} \&
  {Ponman}}{2003}]{Read03}
{Read} A.~M.,  {Ponman} T.~J.,  2003, \mn@doi [\aap]
  {10.1051/0004-6361:20031099}, \href
  {https://ui.adsabs.harvard.edu/abs/2003A&A...409..395R} {409, 395}

\bibitem[\protect\citeauthoryear{{Rines}, {Mahdavi}, {Geller}, {Diaferio},
  {Mohr}  \& {Wegner}}{{Rines} et~al.}{2001}]{2001ApJ...555..558R}
{Rines} K.,  {Mahdavi} A.,  {Geller} M.~J.,  {Diaferio} A.,  {Mohr} J.~J.,
  {Wegner} G.,  2001, \mn@doi [\apj] {10.1086/321513}, \href
  {https://ui.adsabs.harvard.edu/abs/2001ApJ...555..558R} {555, 558}

\bibitem[\protect\citeauthoryear{{Roediger}, {Br{\"u}ggen}, {Simionescu},
  {B{\"o}hringer}, {Churazov}  \& {Forman}}{{Roediger}
  et~al.}{2011}]{Roediger2011}
{Roediger} E.,  {Br{\"u}ggen} M.,  {Simionescu} A.,  {B{\"o}hringer} H.,
  {Churazov} E.,   {Forman} W.~R.,  2011, \mn@doi [\mnras]
  {10.1111/j.1365-2966.2011.18279.x}, \href
  {http://adsabs.harvard.edu/abs/2011MNRAS.413.2057R} {413, 2057}

\bibitem[\protect\citeauthoryear{{Roediger}, {Lovisari}, {Dupke}, {Ghizzardi},
  {Br{\"u}ggen}, {Kraft}  \& {Machacek}}{{Roediger}
  et~al.}{2012}]{Roediger2012b}
{Roediger} E.,  {Lovisari} L.,  {Dupke} R.,  {Ghizzardi} S.,  {Br{\"u}ggen} M.,
   {Kraft} R.~P.,   {Machacek} M.~E.,  2012, \mn@doi [\mnras]
  {10.1111/j.1365-2966.2011.20287.x}, \href
  {http://adsabs.harvard.edu/abs/2012MNRAS.420.3632R} {420, 3632}

\bibitem[\protect\citeauthoryear{{Rossetti}, {Eckert}, {De Grandi},
  {Gastaldello}, {Ghizzardi}, {Roediger}  \& {Molendi}}{{Rossetti}
  et~al.}{2013}]{Rossetti2013}
{Rossetti} M.,  {Eckert} D.,  {De Grandi} S.,  {Gastaldello} F.,  {Ghizzardi}
  S.,  {Roediger} E.,   {Molendi} S.,  2013, \mn@doi [\aap]
  {10.1051/0004-6361/201321319}, \href
  {http://adsabs.harvard.edu/abs/2013A%26A...556A..44R} {556, A44}

\bibitem[\protect\citeauthoryear{{Ruggiero} \& {Lima Neto}}{{Ruggiero} \& {Lima
  Neto}}{2017}]{Ruggiero2017}
{Ruggiero} R.,  {Lima Neto} G.~B.,  2017, \mn@doi [\mnras]
  {10.1093/mnras/stx744}, \href
  {http://adsabs.harvard.edu/abs/2017MNRAS.468.4107R} {468, 4107}

\bibitem[\protect\citeauthoryear{{Sanders} \& {Fabian}}{{Sanders} \&
  {Fabian}}{2006}]{2006MNRAS.371L..65S}
{Sanders} J.~S.,  {Fabian} A.~C.,  2006, \mn@doi [\mnras]
  {10.1111/j.1745-3933.2006.00209.x}, \href
  {https://ui.adsabs.harvard.edu/abs/2006MNRAS.371L..65S} {371, L65}

\bibitem[\protect\citeauthoryear{Scrucca, Fop, Murphy  \& Raftery}{Scrucca
  et~al.}{2016}]{Scrucca16}
Scrucca L.,  Fop M.,  Murphy T.~B.,   Raftery A.~E.,  2016, \mn@doi [{The R
  Journal}] {10.32614/RJ-2016-021}, 8, 289

\bibitem[\protect\citeauthoryear{{Sheardown} et~al.,}{{Sheardown}
  et~al.}{2018}]{Sheardown2018}
{Sheardown} A.,  et~al., 2018, \mn@doi [\apj] {10.3847/1538-4357/aadc0f}, \href
  {https://ui.adsabs.harvard.edu/abs/2018ApJ...865..118S} {865, 118}

\bibitem[\protect\citeauthoryear{{Springel}}{{Springel}}{2005}]{Springel2005}
{Springel} V.,  2005, \mn@doi [\mnras] {10.1111/j.1365-2966.2005.09655.x},
  \href {http://adsabs.harvard.edu/abs/2005MNRAS.364.1105S} {364, 1105}

\bibitem[\protect\citeauthoryear{{Ueda}, {Kitayama}  \& {Dotani}}{{Ueda}
  et~al.}{2017}]{Ueda2017}
{Ueda} S.,  {Kitayama} T.,   {Dotani} T.,  2017, \mn@doi [\apj]
  {10.3847/1538-4357/aa5c3e}, \href
  {https://ui.adsabs.harvard.edu/abs/2017ApJ...837...34U} {837, 34}

\bibitem[\protect\citeauthoryear{{Ueda}, {Ichinohe}, {Kitayama}  \&
  {Umetsu}}{{Ueda} et~al.}{2019}]{Ueda2019}
{Ueda} S.,  {Ichinohe} Y.,  {Kitayama} T.,   {Umetsu} K.,  2019, \mn@doi [\apj]
  {10.3847/1538-4357/aafa19}, \href
  {https://ui.adsabs.harvard.edu/abs/2019ApJ...871..207U} {871, 207}

\bibitem[\protect\citeauthoryear{{Walker}, {Fabian}  \& {Sanders}}{{Walker}
  et~al.}{2014}]{Walker2014}
{Walker} S.~A.,  {Fabian} A.~C.,   {Sanders} J.~S.,  2014, \mn@doi [\mnras]
  {10.1093/mnrasl/slu040}, \href
  {http://adsabs.harvard.edu/abs/2014MNRAS.441L..31W} {441, L31}

\bibitem[\protect\citeauthoryear{{Walker}, {ZuHone}, {Fabian}  \&
  {Sanders}}{{Walker} et~al.}{2018}]{Walker2018}
{Walker} S.~A.,  {ZuHone} J.,  {Fabian} A.,   {Sanders} J.,  2018, \mn@doi
  [Nature Astronomy] {10.1038/s41550-018-0401-8}, \href
  {https://ui.adsabs.harvard.edu/abs/2018NatAs...2..292W} {2, 292}

\bibitem[\protect\citeauthoryear{{Walker}, {Mirakhor}, {ZuHone}, {Sanders},
  {Fabian}  \& {Diwanji}}{{Walker} et~al.}{2022}]{Walker2022}
{Walker} S.~A.,  {Mirakhor} M.~S.,  {ZuHone} J.,  {Sanders} J.~S.,  {Fabian}
  A.~C.,   {Diwanji} P.,  2022, \mn@doi [\apj] {10.3847/1538-4357/ac5894},
  \href {https://ui.adsabs.harvard.edu/abs/2022ApJ...929...37W} {929, 37}

\bibitem[\protect\citeauthoryear{{Yu}, {Serra}, {Diaferio}  \& {Baldi}}{{Yu}
  et~al.}{2015}]{Yu15}
{Yu} H.,  {Serra} A.~L.,  {Diaferio} A.,   {Baldi} M.,  2015, \mn@doi [\apj]
  {10.1088/0004-637X/810/1/37}, \href
  {https://ui.adsabs.harvard.edu/abs/2015ApJ...810...37Y} {810, 37}

\bibitem[\protect\citeauthoryear{{ZuHone}}{{ZuHone}}{2011}]{ZuHone2011}
{ZuHone} J.~A.,  2011, \mn@doi [\apj] {10.1088/0004-637X/728/1/54}, \href
  {http://adsabs.harvard.edu/abs/2011ApJ...728...54Z} {728, 54}

\bibitem[\protect\citeauthoryear{{ZuHone}, {Markevitch}  \& {Johnson}}{{ZuHone}
  et~al.}{2010}]{ZuHone2010}
{ZuHone} J.~A.,  {Markevitch} M.,   {Johnson} R.~E.,  2010, \mn@doi [\apj]
  {10.1088/0004-637X/717/2/908}, \href
  {http://adsabs.harvard.edu/abs/2010ApJ...717..908Z} {717, 908}

\bibitem[\protect\citeauthoryear{{van Haarlem} \& {van de Weygaert}}{{van
  Haarlem} \& {van de Weygaert}}{1993}]{vanHaarlem93}
{van Haarlem} M.,  {van de Weygaert} R.,  1993, \mn@doi [\apj]
  {10.1086/173416}, \href
  {https://ui.adsabs.harvard.edu/abs/1993ApJ...418..544V} {418, 544}

\makeatother
\end{thebibliography}

\label{lastpage}

\end{document}